%% file: main.tex
\documentclass[
authoryear,
preprint,times,review,longtitle,3p
]{elsarticle}
\usepackage[T1]{fontenc}
\usepackage[publisher=elsevier]{arXiv-notices}

\usepackage{amssymb}
\usepackage{pdfpages}

\usepackage{lineno}

\usepackage{rotating} 
\usepackage[colorlinks]{hyperref}
\usepackage{longtable}
\usepackage{array}
\usepackage{soul}
\usepackage{float}
\usepackage{calc}
\usepackage{multirow}
\usepackage{booktabs}
\usepackage{makecell}
\usepackage{graphicx}
\usepackage{caption}
\usepackage{subcaption}
\usepackage[capitalise]{cleveref}
\usepackage{rotating}
\usepackage[nolist]{acronym}
\usepackage[inline]{enumitem}
\usepackage{lscape}

\makeatletter
\def\paragraph{\secdef{\els@aparagraph}{\els@bparagraph}}
\def\els@aparagraph[#1]#2{\elsparagraph[#1]{#2.}}
\def\els@bparagraph#1{\elsparagraph*{#1.}}

\renewcommand\elsparagraph{\@startsection{paragraph}{4}{\parindent}%
           {0pt}%
           {-6\p@}%
           {\normalfont\itshape}}
\makeatother

\usepackage[horizontal,grid=lightgray,skipempty]{credits}

\journal{Int. J. Hum.-Comp. Studies}

\usepackage{changes}

\definechangesauthor[color=red,name=Andrea Esposito]{AE}
\definechangesauthor[color=blue,name=Giuseppe Desolda]{GD}
\definechangesauthor[color=green,name=Francesco Greco]{FG}
\definechangesauthor[color=orange,name=Cesare Tucci]{CT}

\begin{document}

\begin{frontmatter}


\title{Understanding User Mental Models in AI-Driven Code Completion Tools: Insights from an Elicitation Study}

\arXivNotice

\author[uniba]{Giuseppe Desolda\corref{cor1}}
\ead{giuseppe.desolda@uniba.it}
\ead[orcid]{https://orcid.org/0000-0001-9894-2116}
\cortext[cor1]{Corresponding author}
\author[uniba]{Andrea Esposito}
\ead{andrea.esposito@uniba.it}
\ead[orcid]{https://orcid.org/0000-0002-9536-3087}
\author[uniba]{Francesco Greco}
\ead{francesco.greco@uniba.it}
\ead[orcid]{https://orcid.org/0000-0003-2730-7697}
\author[uniba,unisa]{Cesare Tucci}
\ead{ctucci@unisa.it}
\ead[orcid]{https://orcid.org/0000-0001-5181-7115}
\author[uniba]{Paolo Buono}
\ead{paolo.buono@uniba.it}
\ead[orcid]{https://orcid.org/0000-0002-1421-3686}
\author[uniba]{Antonio Piccinno}
\ead{antonio.piccinno@uniba.it}
\ead[orcid]{https://orcid.org/0000-0003-1561-70736}

\affiliation[uniba]{%
    organization={Department of Computer Science, University of Bari Aldo Moro},
    addressline={Via E. Orabona 4},
    city={Bari},
    postcode={70125},
    country={Italy}
}
\affiliation[unisa]{%
    organization={University of Salerno},
    addressline={Via Giovanni Paolo II 132},
    city={Fisciano (Salerno)},
    postcode={84084},
    country={Italy}
}

\begin{abstract}
Integrated Development Environments increasingly implement AI-powered code completion tools (CCTs), which promise to enhance developer efficiency, accuracy, and productivity. However, interaction challenges with CCTs persist, mainly due to mismatches between developers' mental models and the unpredictable behavior of AI-generated suggestions, which is an aspect underexplored in the literature. We conducted an elicitation study with 56 developers using co-design workshops to elicit their mental models when interacting with CCTs. Different important findings that might drive the interaction design with CCTs emerged. For example, developers expressed diverse preferences on when and how code suggestions should be triggered (proactive, manual, hybrid), where and how they are displayed (inline, sidebar, popup, chatbot), as well as the level of detail. It also emerged that developers need to be supported by customization of activation timing, display modality, suggestion granularity, and explanation content, to better fit the CCT to their preferences. To demonstrate the feasibility of these and the other guidelines that emerged during the study, we developed ATHENA, a proof-of-concept CCT that dynamically adapts to developers' coding preferences and environments, ensuring seamless integration into diverse workflows.
\end{abstract}


\begin{highlights}
    \item Developers expressed diverse preferences on when and how code suggestions should be triggered; this highlights the need for both user control and context-sensitive automation.
    \item The position to visualize code suggestion and explanation can be inline for short completions; sidebar, pop-up, or chatbot for longer or more complex suggestions.
    \item Most users want to receive minimal code suggestions and then request details on demand, including support for varying from code statements to whole functions or files.
    \item Explanations should be concise, contextual, and user-triggered (via click or shortcut), with links to documentation or example code.
    \item Developers prefer to customize activation timing, visualization style, explanation detail, and code style according to their personal preferences and project needs.
\end{highlights}


\end{frontmatter}


\begin{acronym}
    \acro{IDE}{Integrated Development Environment}
    \acro{AI}{Artificial Intelligence}
    \acro{CCT}{Code Completion Tool}
    \acro{HCAI}{Human-Centered Artificial Intelligence}
    \acro{HCI}{Human-Computer Interaction}
    \acro{LLM}{Large Language Model}
    \acro{HCI}{Human-Computer Interaction}
    \acro{CNN}{Convolutional Neural Network}
    \acro{XAI}{eXplainable AI}
    \acro{ML}{Machine Learning}
    \acro{UI}{User Interface}
\end{acronym}

\section{Introduction}\label{introduction}

\acp{IDE} support developers with valuable tools to write and debug code in current software development practices. In many \acp{IDE}, code suggestion by automated systems has become a key element. Employing ever-developing \ac{AI} capabilities, these \emph{\acp{CCT}} intend to boost developer efficiency, accuracy, and productivity by cutting down on the manual input needed for repetitive code segments and speeding up the entire coding task \citep{Weber2024Significant,Cui2024Effects,Peng2023Impact}. \acp{CCT} powered by \ac{AI}, like GitHub Copilot and JetBrains \ac{AI}, indicate a movement towards advanced and informed programming assistance.

When dealing with \acp{CCT}, while potentially benefitting from increased efficiency, users are exposed to a wide range of risks strictly related to automation. For example, one of the main risks in the domain of code completion is deskilling \citep{Sambasivan2022Deskilling}, i.e., the risk of losing personal abilities in the automated task (in this case, code writing). Similarly, and as already highlighted by previous studies \citep{Sergeyuk2025Using}, \acp{CCT} may increase the risk of generating unsecured code. In response to the challenges that generally affect interaction with \ac{AI} systems, as in the case of \acp{CCT}, a new field of study has emerged in recent years, lying at the intersection of \ac{HCI} and \ac{AI}, commonly known as \ac{HCAI} \citep{Shneiderman2022HumanCentered,Xu2019HumanCentered}. \ac{HCAI} proposes to develop systems that are designed, developed, and evaluated by involving users in the process, aiming to increase performances and satisfaction in specified tasks \citep{Desolda2024Humancentred}.

While the technical aspects of \ac{AI}'s role in code generation are well-researched (for example, algorithms and AI models used by such tools \citep{izadi2024language, husein2024large}), \ac{HCAI} systems for code completions are still underresearched, exposing developers (and their software) to various risks. Some recent studies confirm these symptoms: only 40\% of CCT suggestions are accepted the first time they are proposed to developers, mainly because developers cannot control context or granularity \citep{Liang2024Survey}. Other evidence about the misalignment between developers' mental models and CCTs regards the developers' need for finer-grained suggestions and trustworthy explanations \citep{Sergeyuk2025Using,Wang2023How}, typically not provided in CCTs, as well as the different preferences about inline and sidebar layouts  \citep{Vaithilingam2023IntelliCode, Marasoiu2015CodeCompletion}. Although some studies explored coders' expectations when dealing with \acp{CCT} \citep{Sergeyuk2025Using,Wang2023How}, the understanding of the cognitive and interaction dynamics that developers face upon adopting \acp{CCT} is limited. To the best of our knowledge, no existing studies empirically map the mental models of developers on the interaction with CCTs, models that might determine when code suggestions and explanations are trusted, understood, or disrupted. This aspect limits both theoretical understanding and practical guidance for interaction designers of CCTs. Understanding developers' mental models can thus lead to developing an \acp{IDE} that aligns with user requirements, fosters user satisfaction, and improves coding productivity while promoting their adoption \citep{Sergeyuk2025Using}.

This study addresses the lack of understanding of the user's mental model in the usage of \acp{CCT} by reporting an elicitation study carried out as 8 co-design workshops with a total of 56 participants. Driven by the limitations identified during the analysis of the state of the art, the study aims to answer the following research questions: 
\begin{enumerate}[label={RQ\arabic*:},ref={RQ\arabic*},leftmargin=*]
    \item How do developers mentally model their interaction with AI-driven \acp{CCT}?
    \item What are developers’ expectations regarding the delivery of explanations alongside AI-generated code suggestions?
    \item How do developers perceive and articulate their needs for personalization in \acp{CCT}?
    \item What human-centered design principles emerge for aligning \acp{CCT} with developer mental models?
\end{enumerate}
Insights gained from this analysis will help dictate strategies for creating user-friendly interaction approaches in \acp{IDE} while increasing cohesiveness among developers and \ac{AI} technologies in software development. 

\subsection{Contributions}
This work advances the understanding and design of AI-driven Code Completion Tools (CCTs) from a Human-Centered AI perspective by making the following key contributions:
\begin{itemize}
    \item\textit{Exploratory Survey of CCTs:} An exploratory survey has been conducted to identify and analyze AI-assisted CCTs, their interaction modalities, and explanation features. We contributed in this way by establishing the landscape of available approaches. Moreover, this activity informed the design of our elicitation study, ensuring that workshops prompts addressed both prevalent and underexplored aspects of developers interacting with CCTs.
    \item \textit{Empirical mapping of developer mental models}: we contribute to the state of the art by proposing the first qualitative elicitation of mental models of CCTs, to explore how developers conceptualize, interact with, and personalize CCTs. Our analysis revealed the multifaceted preferences regarding when code suggestions should appear (manual, proactive, hybrid), where and how they should be displayed (inline, sidebar, popup, chatbot), the granularity and content of suggestions, and the need for contextual, purpose-driven explanations.
    \item \textit{Actionable design guidelines for human-centered CCTs}: study findings have been distilled into concrete design guidelines, which emphasize that no single interaction model can fit all users or tasks. Indeed, users need highly customizable and adaptive code and explanation suggestions, including per-user control over activation mode, visualization, suggestion scope, explanation detail, and integration with individual and project coding styles.
    \item \textit{Prototype demonstration of personalizable CCTs}: a prototype of CCT called ATHENA has been developed to demonstrate the feasibility of the proposed guidelines, as well as to provide a platform for future comparative studies on developer-AI co-adaptation.
\end{itemize}

Together, these contributions provide a significant contribution to the HCI literature on AI development tools, providing evidence and solutions for designing trustworthy, effective, and personally relevant CCTs.

\subsection{Organization}
The manuscript is structured as follows. \Cref{rationale-and-background} presents the rationale and background of our study, providing a survey of existing \acp{CCT} and their interaction modalities. \Cref{methodology} presents the methodology of our study, detailing the co-design workshops. \Cref{findings} reports the findings of our elicitation study. \Cref{discussion} discusses the findings, presenting developers' preferences and their mental model of \emph{human-centered} \ac{AI}-based \acp{CCT}. \Cref{prototype-development} illustrates a first working prototype that implements the elicited mental model. \Cref{threats} discusses the elements of the study that could threaten its validity and describes the measures taken to mitigate them. Finally, \Cref{conclusion} concludes the article, presenting its limitations and potential future work.

\section{Rationale and Background}\label{rationale-and-background}

\ac{AI}-driven \acp{CCT} are influencing developers' approach toward programming tasks \citep{Alenezi2025AIDriven,Martinovic2024Impact}. Real-time assistance is provided through these tools with suggestions for code snippets, auto-completion of code syntax, and predicting coding patterns while improving productivity, lowering the cognitive load, and streamlining the coding process. Although the implementation and usability of \acp{CCT} very much rely on the developer's expectations, understanding, and workflows, the success of these tools largely relies on how closely they align with what developers are already using, what they expect to use, their skills, and their needs. Achieving this alignment mainly hinges on users' mental models, i.e., internal representations of how the system works.

\subsection{Human-centered AI and code completion tools}

The rapid growth of \ac{AI} in recent years has significantly increased its adoption and impact in research and business activities. However, AI's pervasiveness has exacerbated concerns over existing flaws. In particular, biases and lack of explainability endanger the users of the AI models, which often do not consider the human element \citep{Shneiderman2020Bridging}. A report by the \citet{NationalTransportationSafetyBoard2017Collision} regarding a deadly crash of an autonomous Tesla car stated the following: <<automation ``because we can'' does not necessarily make the human-automation system work better. [...] This crash is an example of what can happen when automation is introduced “because we can”  without adequate consideration of the human element>>. To mitigate these issues, the novel field of \ac{HCAI} is gaining traction \citep{Shneiderman2022HumanCentered}. One of the main goals of \ac{HCAI} is to produce \ac{AI} systems that are Reliable, Safe, and Trustworthy \citep{Shneiderman2020HumanCentered}. For this purpose, AI systems should provide a high level of computer automation while guaranteeing a high level of human control when desired---i.e., AI systems should not automate tasks but amplify, augment, empower, and enhance their users' skills.

In the context of \acp{CCT}, systems can be designed either by aiming at \emph{full automation} or at \emph{full augmentation} (or anywhere in between). On one hand, a full-automation approach can provide systems that are extremely efficient, they are associated with several risks. In addition to an increase in commission and omission errors \citep{Cummings2004Automation}, full-automation \acp{CCT} increase the risk of deskilling \citep{Sambasivan2022Deskilling}. On the other hand, previous user studies highlighted that using a full-augmentation approach ``by default'' may lead to systems that are not on par with users' expectations, thus making them inefficient and ineffective \citep{Esposito2024Fine}.
In this context, our study emphasizes the need to design for augmentation, while still aligning \acp{CCT} with users’ mental models, thus ensuring that developers retain full control over their tools and allowing them to balance the two approaches.

\subsection{Usability studies on code completion tools}\label{studies-in-the-field-of-hci-on-code-completion-tools}

The usability of \acp{CCT} has been a research focus for quite some time, aiming to make these tools more intuitive and helpful for developers. One of the first attempts was carried out by \citet{muaruașoiu2015empirical}, which empirically investigated how professional developers interact with \acp{CCT}. They analyzed the behaviors, intentions, and obstacles programmers encounter when employing these instruments. The study results revealed that code completion is primarily employed to speed up the coding process and guarantee accuracy. Besides, developers frequently use it as real-time feedback to identify mistakes.

Another aspect investigated by HCI researchers focused on the users' behavior while adopting \acp{CCT}. From this study, two main modalities emerged \citep{Prather2024Its}: in ``acceleration'' mode, the programmer is aware of their next steps and utilizes the tool to speed up the coding process. In contrast, in the ``exploration'' mode, the programmer is uncertain about how to proceed and employs the tool to investigate his options. Acceleration mode is usually preferred, as developers often use the tools to complete repetitive code that cannot be copy-pasted \citep{Liang2024LargeScale}, and for recalling syntax they don't remember \citep{Xu2022IDE}. As a result, the more developers accept \acp{CCT} suggestions, the more they perceive themselves as productive \citep{Bird2022Taking}.

\citet{Liang2024LargeScale} conducted a study to understand the usability of \acp{CCT} (such as GitHub Copilot and Tabnine) by utilizing a structured survey distributed to 410 GitHub users who had engaged with \acp{CCT}. Questions covered participants' motivations for using or not using \ac{AI} assistants, the usability issues they encountered, and the strategies they used to optimize the tool's output. The survey revealed that developers are most motivated to use AI programming assistants because they help developers reduce keystrokes, finish programming tasks quickly, and recall syntax. They also found that developers often do not use CCTs because these tools do not generate code that addresses certain functional or non-functional requirements, and because developers have trouble controlling the tool to generate the desired output.

Another important requirement that emerged from user studies is the need for \acp{CCT} to learn from feedback and adapt their behavior to developers' styles and project needs. For instance, \citet{Zhang2023Demystifying} highlight that programmers would like to customize the shortcuts and to set suggestions length and frequency. The literature also highlights the need for explanations (such as reporting the source or linking to documentation) for additional context on the generated code \citep{Liang2024LargeScale,Xu2022IDE}.

The proactivity of \acp{CCT} has also been investigated, with research encouraging its implementation \citep{Sergeyuk2024IDE}. \citet{Vaithilingam2023More} suggested that automatic inline gray-text suggestions, as opposed to traditional lightbulb interfaces, significantly improve discoverability and usability of \acp{CCT}. The authors established five design principles to guide future \ac{AI}-driven code suggestion interfaces:

\begin{itemize}
    \item \textit{Glanceability}. Suggestions should be visible at a glance without manual initiation.
    \item \textit{Juxtaposition}. Clearly displaying original vs. suggested code enhances comprehension.
    \item \textit{Familiarity}. Using familiar visual elements, like red and green colors for highlighting differences with previous code, simplifies understanding.
    \item \textit{Visibility for Validation}. Users should be able to view suggestions fully and evaluate them without additional steps.
    \item \textit{Snoozability}. The ability to temporarily dismiss suggestions reduces interruptions.
\end{itemize}

Many users have asked for a ``snooze'' feature that allows them to pause inline suggestions for a specified period or the current session. This points out the need for further research to find the right balance between proactiveness and minimizing interruptions, ensuring that suggestions are presented at the most helpful moments without disrupting the coding process.

\subsection{The importance of mental models}

A mental model is generally a mental picture users have of how a certain system works \citep{Norman1983Observations,Norman2013Design}. It includes a user's assumptions, beliefs, and expectations while interacting with a system that will lead to their actions, interactions, and problem-solving behavior. Mental models also change over time, might contain errors, and are constrained by the user's prior experiences or technical background. Multiple mental models of a single object may be held by one person to reflect its many functions, and two persons will not always have the same mental model. If mental models are accurate and considered in the design of a system, users can anticipate system responses, feel confident during the interaction, and adapt quickly to new features. On the other hand, gaps between the users' mental model of the system's functionality and its actual behavior can lead users to confusion, frustration, and errors, which will reduce their productivity and satisfaction \citep{190/sym130.33050795}.

\acp{CCT} supported by \ac{AI} employ \ac{ML} algorithms or \acp{LLM} and rely on probabilistic computations rather than being deterministic. For example, \ac{AI}-driven suggestions may use aggregated pattern data, which are not directly readable by users. Thus, the suggestions produced might be contextually nonrelevant or impossible to interpret. If a developer's mental model does not match the underlying mechanisms of an \ac{AI} tool, users might struggle to trust the tool, interpret its suggestions accurately, or recognize when and how to use it effectively.

The need to understand mental models in the domain of \ac{AI}-based \acp{CCT} is multifaceted, and different aspects can benefit from a deep understanding of users' mental models. The first is \emph{interaction} with \acp{CCT} since users might predict the tool's behavior and know when a suggestion will be useful and when it could mislead. For instance, a tool designed considering users' mental models might allow developers to know when the \ac{AI} might make contextually relevant suggestions and when it can blunder, and so on, to keep the coding flow in hand and avoid wasting time for distraction. Another aspect that can benefit regards \emph{trust}. \acp{CCT} can be largely obscure to those operating outside the normal programming logic. It has been proven that transparency and interpretability are essential for establishing user trust, and understanding the user's mental model is essential to making users comfortable when adopting the tools \citep{RN1500}. This helps create trust in a tool by aligning design features with users' mental models, making developers more likely to adopt it into their daily workflow. Another aspect that can benefit from the mental model is \emph{cognitive load}: \acp{CCT} are meant to reduce repetitive tasks and cognitive load; thus, users can concentrate on more difficult problem-solving jobs. However, if users cannot intuitively understand how or why these suggestions are being made, the cognitive load can actually rise rather than fall. Refining these tools to become less disjointed in relation to user mental models means we can improve the flow of work, minimize interruptions, and ultimately help users be more productive. The last aspect is the possible \textit{support to learning}: especially for novice programmers, developers can recognize common coding patterns and problem-solving techniques, enhancing both tool competency and coding proficiency. The overall improvements in systems that stem from following users' mental models throughout the design are also a potent driver for increasing systems' adoption and acceptability \citep{Sergeyuk2025Using}. A recent study by \citet{Wang2023How} highlights that 54\% of the developers recruited for their study are eager for better \acp{CCT}.

Exploring the users' mental models is a complex process since these models are typically abstract, partially formed, or even subconscious. Without elicitation studies, bridging this gap is an arduous task, and they are the ones that prescribe structured methodologies to help yield, analyze, and interpret users' cognitive models regarding system functionality. The techniques for eliciting mental models include design workshops \citep{Jacko2012Human}, interviews \citep{Rogers2023Interaction}, concept mapping \citep{Jacko2012Human}, think-aloud protocols \citep{Lewis1982Using}, and scenario-based design \citep{Rosson2012ScenarioBased}. These techniques allow researchers to design intricate representations of users' mental models of system behavior, what they expect the system to do \emph{a priori}, and how they can expect to interact with, for example, \acp{CCT}. Such insights inform tool design and can help bolster our understanding of \ac{HCI} within \ac{AI}-augmented coding domains towards supporting frameworks and guidelines for improved user experiences in such environments and beyond.

\section{Exploratory survey of AI code completion tools}\label{ai-code-completion-tools}

To inform the design of our elicitation study, we conducted an \textit{exploratory survey} of innovative AI-assisted CCTs. Rather than aiming for comprehensive or systematic coverage, this survey aims to identify tools that are either widely adopted in the industry or that are characterized by distinctive interaction modalities, relevant to our research questions. We identified candidate CCTs through a review of recent academic publications, developer forums, official product websites, and prominent marketplace listings (e.g., Visual Studio Marketplace, GitHub repositories) as of April 2024. CCTs were included if they provided a graphical user interface and exposed features for code suggestion and/or explanation. We focused on selecting those CCTs that represented the breadth of current interface and explanation approaches encountered by developers in practice.

The principal goal was not to exhaustively catalogue the CCT landscape, but to develop a functional mapping of interaction paradigms and explanation methods. This mapping directly informed the design of our empirical study by grounding scenario construction and workshop prompts in real, contemporary CCT examples. While acknowledging limitations inherent in a non-systematic approach, this selection ensures our study addresses key facets of developer–CCT interaction as manifested in actual products.

\subsection{Code completion tools analysis}
The proliferation of powerful \ac{AI} tools, such as \acp{LLM}, is stimulating the creation of many \acp{CCT}. The analysis of the literature and the market allowed us to identify a total of 17 tools. Each tool is characterized by many aspects, ranging from technical features (e.g., the underlying \ac{AI} model) to more social aspects, such as interaction with them. Given the purpose of this study, we analyzed the 17 \acp{CCT} with respect to the interaction techniques offered. However, 4 do not provide a graphical user interface, so they are not considered in the following.

The analysis of the remaining 13 \acp{CCT} providing a \ac{UI} was guided by two main aspects: the \emph{code suggestion} itself and the \emph{explanation}, if any, of why that code was suggested. After familiarizing ourselves with the \acp{CCT}, we analyzed users' interaction with them across different dimensions. To simplify the analysis and the reporting of the results, we used the 5W model by \citet{Lasswell1974Structure}, which is adopted in various fields, such as journalism and customer analysis, and more generally in problem-solving, to analyze the complete story of a fact. A similar approach is employed in the field of \ac{HCI}, where the 5W model is utilized for mental model elicitation, and comprehensive user analysis \citep{Jacko2012Human,Desolda2017Empowering,Duric2002Integrating}. Each dimension of the interaction with the tool was addressed in terms of the question of the 5W model. In this phase, the 5W model guides the analysis of the state of the art of \acp{CCT}. The questions used in this phase are similar to those presented in \cref{procedure}, which are used to elicit users' mental model:
this choice allows the comparison of existing \acp{CCT} with users' expectations. This choice allows to leverage existing tools (identified during this analysis) as design inspirations during prototype development (see \cref{prototype-development}), since it allows the potential identification of tools that satisfy users' expectations (elicited in the co-design workshops, see \cref{methodology}). The analysis of the tools with respect to the code suggestion was done by answering these questions:

\begin{enumerate}
    \item \emph{Where} to show the code completion in the \ac{UI}?
    \item \emph{When} should the suggestion be activated during the interaction?
    \item \emph{How} should the suggested code be shown?
    \item \emph{What} should be suggested?
    \item \emph{How} should the suggestion be customized?
\end{enumerate}

\noindent The analysis regarding the explanations was guided by the following questions:

\begin{enumerate}[resume]
    \item \emph{Where} to show the code completion explanation in the UI?
    \item \emph{When} to show the explanation during the interaction?
    \item \emph{What} should be explained?
\end{enumerate}

\noindent A summary of the complete set of such models is depicted in \cref{tab:tools}.

\begin{table*}[t]
\centering
\caption{Summary of the analysis of  interaction with code and explanations generated by 13 Code Completion Tools.}
\label{tab:tools}
\resizebox{\textwidth}{!}{%
\begin{tabular}{|>{\bfseries}l|l|l|l|l|l|l|l|l|}
\hline
\multirow{2}{*}{} & \multicolumn{5}{c|}{\textbf{Code completion}} & \multicolumn{3}{c|}{\textbf{Explanation}} \\ \cline{2-9}
 & \textbf{What} & \textbf{When} & \textbf{How to show} & \textbf{How to customize} & \textbf{Where} & \textbf{What} & \textbf{When}  & \textbf{Where}  \\ \hline
Intellicode~\citep{Svyatkovskiy2020IntelliCode} & lines & proactive & grey text & curent file & inline & - & -  & -  \\ \hline
Codex~\citep{Chen2021Evaluating} & lines & manual & text & chat history & right side & answer & prompt  & chatbot  \\ \hline
Copilot~\citep{Friedman2021Introducing} & lines & proactive & grey text & current file & inline & answer & prompt  & chatbot \\ \hline
TravTrans~\citep{Kim2021Code} & token & proactive & dropdown & current file & inline & - & -  & - \\ \hline
StarCoder~\citep{li2023starcoder} & lines & proactive & grey text & current file & inline & - & -  & -  \\ \hline
Codegeex~\citep{Zheng2023CodeGeeX} & lines & proactive & grey text & current file & inline & answer, description & prompt, click & chatbot \\ \hline
CodeWhisperer~\citep{What} & lines & hybrid & grey text & current file & inline & source & click  & sidebar  \\ \hline
Codeium~\citep{Codeium} & lines & proactive & grey text & current file & inline & improvements, reasons & click  & chatbot \\ \hline
Replit~\citep{Replit} & lines & proactive & grey text & current file & inline & description & click  & pointer  \\ \hline
Cody~\citep{Cody} & lines & proactive & grey text & repository & inline & answer & click, prompt  & chatbot  \\ \hline
Tabnine~\citep{Tabnine} & lines & proactive & grey text & current file & inline & answer & prompt  & chatbot  \\ \hline
Xcode~\citep{Xcode} & lines & proactive & grey text & current file & inline & - & - & -  \\ \hline
Blackbox~\citep{Chat} & lines & proactive & grey text & current file & inline & improvement & click  & sidebar  \\ \hline
\end{tabular}%
}
\end{table*}

\subsection{Code suggestions}\label{code-suggestions}

An important characteristic affecting the interaction with \acp{CCT} is the timing to provide suggestions (\emph{when}). The analysis of the existing tools revealed two main approaches: \emph{proactive} and \emph{manual}. Proactive \acp{CCT} 
provide suggestions at a certain time without requiring a user action to trigger the completion. This type of design can be either implemented in an intrusive manner (by suggesting a completion at any time)  or discreetly (by generating completions only when the context is sufficient). The manual approach, in contrast, requires the user to perform a specific action to visualize the completion. However, manual activation of \acp{CCT} is not very popular, as it is adopted in only two tools. A hybrid timing approach is also implemented, even if only in the case of CodeWhisperer, as completions are proactive but can be forced through a shortcut \citep{What}.

Another aspect is the content of \acp{CCT} suggestions (\emph{what} is provided as a suggestion?). Two main modalities emerged from our analysis: \emph{single token suggestions} (less popular, as it was proposed only with the TravTrans prototype \citep{Kim2021Code}) and \emph{multiple line completions}, which are widely adopted. While the first solution is quite simple, the second one results in a feature that is hard to control and assess since the length of the suggestions may vary depending on the context and \ac{LLM} characteristics. As a matter of fact, multi-line suggestion length can range from short code blocks (e.g., for loops, switches, if-else) to entire functions or classes.

The suggestion's location (\emph{where}) in the \ac{UI} is also worth exploring. A predominant approach is to put completions directly inside the code (\emph{inline suggestions}). The other solution instead consists of providing the suggestion in a \emph{separate space} with respect to the coding area. Codex \citep{Chen2021Evaluating} is the only tool that adopts the latter modality.

It is useful to discuss \emph{how to show} code completions that appear before they are accepted and integrated into the live code. A consolidated approach is to display the completed code in gray, allowing users to easily distinguish it from the live code, which is typically shown in white (or other colors, depending on the users' preferred color scheme), before accepting the proposals of the \ac{CCT}. Another method that emerged from the analysis is displaying multiple possible completions in a dropdown menu, which may be particularly convenient for single-token suggestions or very short completions, as shown by TravTrans.

Finally, it is worth investigating the customizability of \acp{CCT}, as the relevance and helpfulness of suggestions strongly depend on the type of user receiving them (\emph{how to customize}). Context-aware suggestions may consider the user expertise, the specific formatting/adherence to coding standards, and the functional and non-functional requirements of specific software development scenarios to provide the most accurate code snippets possible. The strategies adopted for this aim result in three categories: personalization based on the context of the current file, personalization based on a code repository, and personalization based on chat history. The first category is the most popular, while Cody \citep{Cody} is the only tool that managed to include a code repository for more relevant completions. Nevertheless, this last approach seems to fail for big repositories. Codex  provides more aware suggestions by processing the conversation between the programmer and the chatbot as context.

\subsection{Explanations of the suggested code}\label{explanatory-features}

Some of the analyzed \acp{CCT} 
do not offer any explanatory features. By contrast, the remaining tools use various methods to provide developers with explanations as they write new code.

The location in the UI (\emph{where}) of code explanations is an important aspect of \acp{CCT}, as it influences other interaction characteristics. Three main approaches are observed: explanations are either delivered inside a chatbot, or shown in a sidebar, or finally displayed inside the coding space. Tabnine \citep{Tabnine}, Cody \citep{Cody}, CodeGeeX \citep{Zheng2023CodeGeeX}, Copilot \citep{Friedman2021Introducing}, Codeium \citep{Codeium}, and Codex \citep{Chen2021Evaluating} implement the first approach, providing chat-based explanations. 
In contrast, CodeWhisperer \citep{What} and Blackbox \citep{Chat} report code explanations in a sidebar. In this case, the user is not allowed of submitting prompts for additional details, but such an approach delivers code explanations quicker, as the user does not have to think about a prompt in order to receive explanations.
Finally, Replit \citep{Replit} follows a unique approach, showing explanations inside a tooltip at the mouse pointer's location (which means the explanations appear within the coding space).



As mentioned earlier, the position in which explanations are delivered directly impacts their content (\emph{what}). Chat-based explanations typically offer direct responses to specific user questions. Static explanations, in contrast, can vary widely. For instance, CodeGeeX and Replit  provide descriptive text for code, with Replit defaulting to a short, numbered list format, as too much text would be too intrusive to fit inside the coding space. CodeWhisperer lists sources used for generating completions, and Blackbox and Codeium suggest code improvements, with the latter adding implementation rationales to its suggestions.


 The timing approaches (\emph{when}) are exclusively manual for explanations. Two primary modalities emerge: explanations can be triggered by a mouse click or provided after submitting a prompt (in the case of chat-based explanations). CodeWhisperer, Codeium, Replit, and Blackbox follow the former approach, although multiple clicks are often required to access the explanations. The latter approach is used by Codex, Tabnine, and Copilot. CodeGeeX and Cody adopt a hybrid approach, offering explanations through both methods.

\section{Methodology}\label{methodology}

This section details the methodology that was employed in this study. Leveraging the 5W model by \citet{Lasswell1974Structure}, we conducted co-design workshops that aimed at eliciting users' mental models and expectations for \acp{CCT}.

\subsection{Research design}\label{research-design}

This study investigates the mental model of developers interacting with \acp{CCT}. Based on the results and limitations that emerged during the literature analysis, we address the following research questions:

\begin{enumerate}[label={RQ\arabic*:},ref={RQ\arabic*},leftmargin=*]
    \item How do developers mentally model their interaction with AI-driven \acp{CCT}?\label{RQ1}
    \item What are developers’ expectations regarding the delivery of explanations alongside AI-generated code suggestions?\label{RQ2}
    \item How do developers perceive and articulate their needs for personalization in \acp{CCT}?\label{RQ3}
    \item What human-centered design principles emerge for aligning \acp{CCT} with developer mental models?\label{RQ4}
\end{enumerate}

To answers these questions, we followed an elicitation study methodology conducted through design workshops. We have chosen this technique because, particularly when performed in workshop formats, it facilitates in-depth insights into users' thought processes, preferences, expectations, needs, and perceptions about complex systems (similar to what has been done by \citet{Desolda2017Empowering,Marquardt2012Crossdevice,Voida2005Study}). Moreover, workshops stimulate participants to collaboratively propose, discuss, and model their ideas on code completion tools, offering rich qualitative data suited to the study's objectives. To structure the workshop discussions, we adopted a scenario-based design approach. This solution helps participants provide their answers while referring to a realistic situation \citep{Rosson2012ScenarioBased}.

\subsection{Participants}\label{participants}

\begin{figure*}[t]
    \centering
    \includegraphics[width=\linewidth]{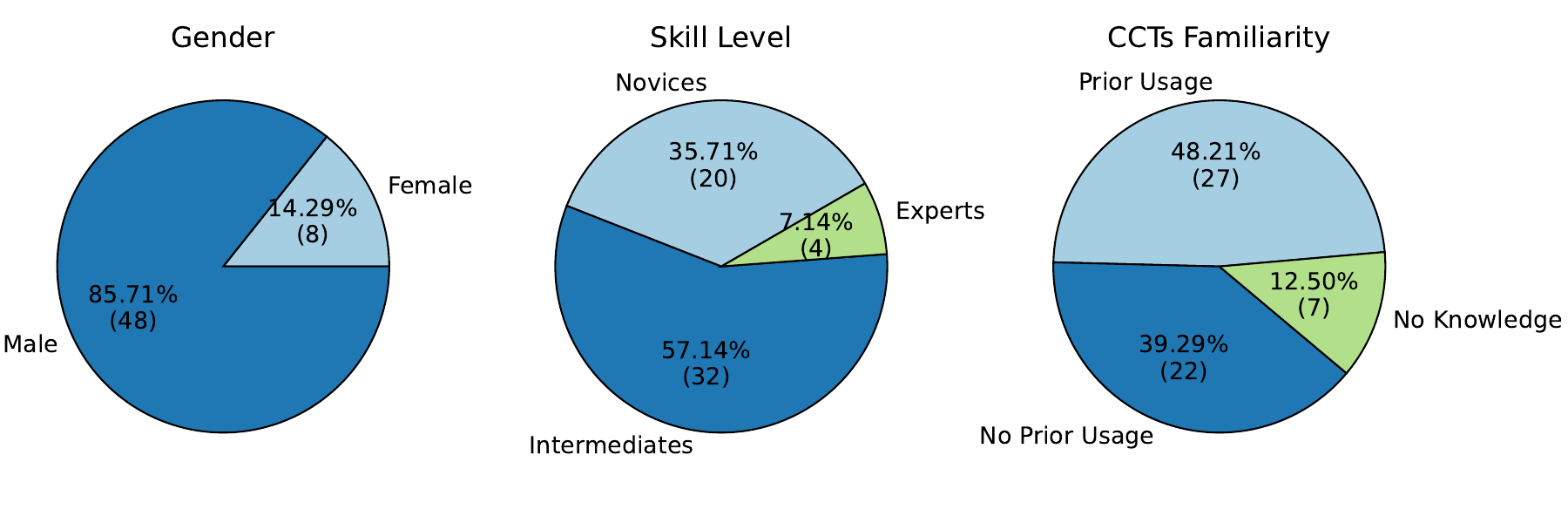}
    \caption{Participants' demographic information}
    \label{fig:participants}
\end{figure*}

The elicitation study involved a total of 56 participants (8 female, 48 male), aged between 20 and 30 years ($M = 22.88$, $SD = 2.63$). They were randomly divided into 8 groups of 7 participants; this group size was chosen to enable balanced discussions where each participant could contribute their perspectives while fostering dynamic group interactions.  \Cref{fig:participants} presents participants' demographic information. The participants were students from both the University of Bari and the University of Salerno; 4 already had bachelor's degrees in computer science, while the remaining were students in the third year of their bachelor's degrees in computer science. Participants included novice (20), intermediate (32), and expert programmers (4). Their programming experience ranged from 1 year to 8 years ($M=3.36$, $SD=2.07$), with familiarity in languages such as C/C++ (most common ones, as 100\% of participants were familiar with them), followed by Javascript (known by 66\% of participants), Python, Java, and C\#. A total of 27 participants used \acp{CCT}; 22 participants had never used \acp{CCT}, but they knew what \acp{CCT} do, while 7 participants had never used and heard about \acp{CCT}.

All relevant ethical guidelines and regulations were followed, and the study received ethical approval from the Independent Ethical Review Board of the Computer Science Department of the University of Salerno. The participants digitally provided their informed consent and could leave the study at any time.

\subsection{Procedure}\label{procedure}

Two weeks before the start of the workshops, 100 emails were sent to the students inviting them to participate in the study. A total of 58 students agreed to attend the study, but then 2 of them missed the workshops. Those who signed up voluntarily filled in a questionnaire asking for personal data (first name, last name, age, gender), and data on technical skills (self-rating programming experience as novice, intermediate, expert), number of years of experience as a programming professional, use of \acp{CCT} (``yes'', ``No, but I know what they are and how they can work'', ``No, and I don't know what they are''), and used programming languages. We also surveyed their availability for the week of the study to schedule their participation according to their preferences.

Two \ac{HCI} researchers performed the study on four consecutive days in a quiet university laboratory; one was the conductor, and the other acted as the observer. The entire study consisted of 8 sessions, one for each group. Each workshop was planned to last around 1 hour and comprised four stages: introduction, scenario exploration, mental model elicitation, and debriefing.

For each session, the group sat around a table and was provided with paper sheets and markers to sketch their proposals. One of the two \ac{HCI} researchers began by welcoming participants and thanking them for attending the study; then, the conductor gave a 5-minute presentation to provide an overview of the study's objectives, ensuring that participants understood the focus on user interaction with \acp{CCT}. Participants were also informed of the confidentiality and voluntary nature of the study.

To allow participants to recall the main concepts of \acp{CCT} (if they already used them) or to familiarize themselves with them, we presented a scenario involving a typical developer named Andrea, employed in a company producing web applications. In this scenario, Andrea has access to new functions of the \ac{IDE} for which an \ac{AI}-based module can suggest code to help him in his work. No concrete ideas or possible solutions were reported in the scenario to avoid bias in the participants' proposals.

After the introduction and the scenario presentation, the workshop's main phase began. The conductor led the discussion, focusing on those critical aspects of the interaction with \acp{CCT} that emerged from the literature review. In particular, for code completion, we posed the following questions:

\begin{enumerate}
    \item Where to show the code completion in the \ac{UI}?
    \item When should the suggestion be activated during the interaction?
    \item How should the suggested code be shown?
    \item What should be suggested?
    \item How should the suggestion be customized?
\end{enumerate}

\noindent For the explanation part, the discussion was guided by the following questions:

\begin{enumerate}
\setcounter{enumi}{5}
    \item Where to show the code completion explanation in the \ac{UI}?
    \item When to show the explanation during the interaction?
    \item What should be explained?
\end{enumerate}

For each question, the participants were asked to reflect individually for about 30 seconds and present the solution they had thought of to the others; afterward, the group was driven by the conductor in a discussion to come up with one or more ideas to answer the specific question.

Once all the questions had been answered, the session ended with a debriefing to summarise the key points raised by the participants and to revise, if necessary, some of their answers in the light of what had emerged throughout the discussion.

\subsection{Data collection}\label{data-collection}

The data collected during the workshops were 1) the notes taken by the researchers, 2) the audio recordings, and 3) the sketches drawn by participants. The two researchers transcribed the notes and audio recordings, and independently rechecked 80\% of the material. The initial reliability value was 70\%; then, the researchers discussed the differences and reached full reliability.

The researchers analyzed the transcripts systematically using an inductive thematic analysis \citep{Braun2006Using}. According to the thematic analysis protocol, the evaluators performed the following 6 steps: familiarizing with the data, generating codes, mapping codes that share similar meanings, reviewing potential themes, defining and naming themes, and producing a report. The eight guiding questions structured the analysis, with the final themes answering each question.

\section{Findings}\label{findings}

This section presents the lessons learned about the user mental model on code completion for both suggestion and explanation, drawn from the elicitation study. The findings are presented with respect to the questions that structured the discussions during the workshops and provide a concrete description of users' thoughts about desirable interaction with code completion and explanation features. Each finding derives from a single theme that emerged from the thematic analysis, which is, in turn, supported by one or more low-level codes. All the themes, codes, and frequencies are reported in \ref{app:themes}.

The names of each theme are presented in italics in the following subsections. Participants' answers are given in double quotes to illustrate how one of the answers led to the identification of a theme. We also specified the number of groups that proposed the solution linked to each theme. It is worth noting that we chose to report the number of groups rather than the number of participants because, within a group, several solutions for each question were initially proposed individually; however, subsequent discussion among group members led to the identification of one or more solutions to answer the question.

\subsection{UI for code completion}\label{ui-for-code-completion}

\begin{figure}[t]
    \centering
    \includegraphics[width=.8\linewidth]{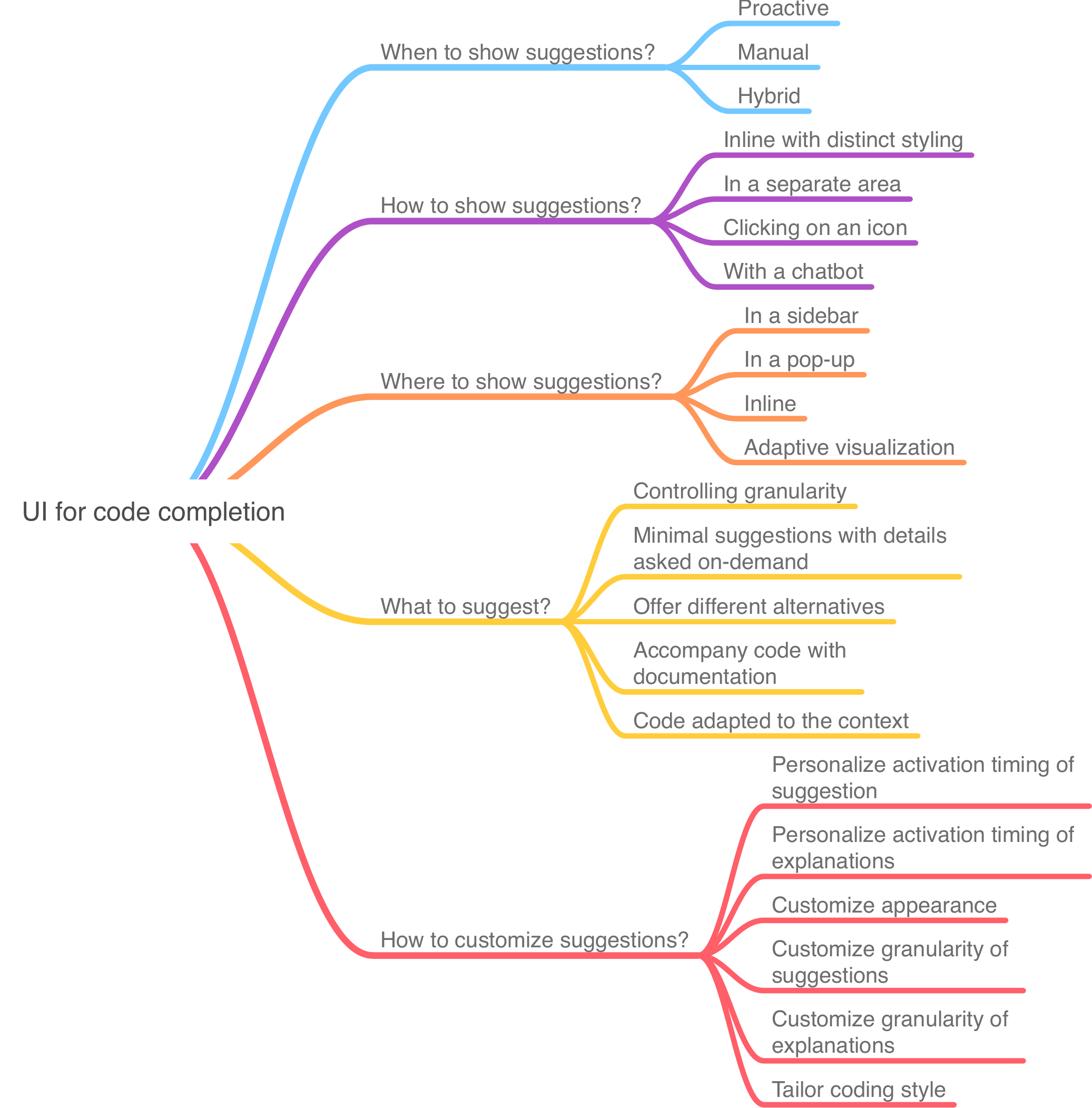}
    \caption{Themes emerged by analyzing the developers' answers to the questions on user interfaces of CCTs.}
    \label{fig:findings-ui}
\end{figure}

The results of this subsection address \ref{RQ1}, since it focuses on how developers mentally model their interaction with AI-driven CCTs. The findings describe developers' preferences on when and how code suggestions should be triggered, displayed, and customized, in order to adhere to users' internal representations and expectations toward CCT behavior. The themes, represented in \cref{fig:findings-ui}, also partially cover \ref{RQ3}, specifically on the developers' needs for customizing activation and display strategies.

\paragraph{When to show suggestions} 
There were differing proposals with regard to when the \acp{CCT} should provide code suggestions for participants. Three main strategies emerged:
\begin{itemize}
    \item \textbf{Proactive} (4 groups): the \ac{CCT} automatically suggests the code without users' requests when the user is not actively typing, e.g., coding `breaks' in the coding flow (``\emph{When someone is blocked, there a suggestion is given to help me}'' -- G1)
    \item \textbf{Manual} (5 groups): activation by keyboard shortcut or mouse interaction (``\emph{It must be activated manually via a shortcut}'' -- G2). This solution leaves complete control to the user over the use of \ac{AI}, which could then be less intrusive.
    \item \textbf{Hybrid} (6 groups): a basic behavior could involve a proactive \ac{AI} strategy that intervenes only when there is a well-defined context (e.g., a long code already written) and when the suggestion is almost taken for granted and difficult to discard by the user (e.g., completion of a control structure). On the other hand, manual intervention would compensate the user for the lack of automatic \ac{AI} intervention, provided that the user is sure that they require \ac{AI} support (``\emph{Proactive when it's really obvious and useful, or when the user explicitly requests it}'' -- G8). Furthermore, several participants specify that, in general, suggestions should be generated only if there is enough \emph{coding context available} (4 groups) (``\emph{It's better not to try immediately, but take from context}'' -- G2). This was also consistent with the perceived necessity of very precise, context-sensitive recommendations \citep{asaduzzaman2014cscc}.
\end{itemize}

\paragraph{How to show suggestions} 
The second aspect concerns how to show the suggestions. Four suggestions emerged:
\begin{itemize}
    \item \textbf{Inline with distinct styling} (5 groups): code suggestions should be displayed within the main coding window, but with differences in formatting, such as unique font styles or colors (``... \emph{chosen the `phantom' code suggested, it is then shown in our code, to give a `preview'}'' -- G6). This method allows users to view suggestions in context while maintaining clarity, particularly when the suggested code is short (``\emph{For little code, it is convenient showing it intext}'' -- G2). In addition, a button could be useful to switch this display on or off, i.e., to show \ac{AI} and user-written code in the same or different fonts (``\emph{... turn on/off the `show AI/human written code' mode...{}}'' -- G8).
    \item \textbf{In a separate area} (7 groups): visualization of suggestions in a separate part from the text editor, such as a sidebar. This modality allows users to review suggestions (especially when they are lengthy or more than one) without cluttering up the main coding space while keeping the focus on the existing code (``\emph{A separate window with a list of suggestions}'' -- G4; ``\emph{For on-demand things or lengthier suggestions, use a separate window}'' -- G3).
    \item \textbf{Clicking on an icon} (2 groups): manually activate suggestions by clicking on a button and displaying it in a popup near the line where the suggestion is applied. By clicking on the icon, users can see and confirm the insertion of the suggestion, minimizing interruptions and increasing the level of control (``\emph{Maybe a `lightbulb' icon that allows opening a popup}'' -- G3). 
    \item \textbf{Using chatbots} (2 groups): allow users to request suggestions in a conversational manner. The suggestion would then be entered directly into the \ac{IDE} when the user accepts the chatbot's recommendation (``\emph{It should be conversational to improve suggestions}'' -- G3).
\end{itemize}

\paragraph{Where to show suggestions}
The third aspect concerns where to display the suggestion in the user interface of the IDE. Discussions with participants led to four strategies: 
\begin{itemize}
    \item \textbf{In a sidebar} (6 groups): the first preference expressed by participants is to show the suggestion in a sidebar (``\emph{It's} \emph{better on the side, separately}'' -- G1; ``\emph{In a movable tab of the IDE}'' -- G7).
    \item \textbf{in a pop-up} (3 groups): the second strategy suggested by participants is to show the suggestion in a pop-up near the main code (``\emph{In a togglable pop-up near the main code}'' -- G3).
    \item \textbf{Inline}(5 groups): another strategy that participants came up with is to show the suggestion inline with the code that the developer is writing; in this case, the user could also be given a clear and immediate option to accept or reject the suggestion, e.g., an accept or reject icon (``\emph{{[}the suggestion{]} is shown within the editor, with the possibility to confirm the insertion'' --} G8).
    \item \textbf{Adaptive visualization} (2 groups): the last mode combines the previous ones in which shorter suggestions appear inline and longer ones are presented in a separate window (``\emph{In loco or in a different area. The choice depends on the code quantity''} -- G2). This approach allows flexibility depending on the length and complexity of the suggestion.
\end{itemize}

\paragraph{What to suggest}
The fourth aspect concerns what to show in the suggestion. Participants expressed different expectations of what constitutes valuable content in code suggestions:
\begin{itemize}
    \item \textbf{Suggestions of various granularity} (7 groups): in general, the strategy of controlling granularity emerged. Notably, participants expressed very contrasting needs for different levels of granularity for code suggestions: while some groups agreed on preferring suggestions up to a maximum granularity level of \emph{function} (``\emph{Limit to completing the function: if it completes everything, randomness increases and precision decreases}'' -- G1), others agreed on the usefulness of generating up to entire code files or classes (``\emph{Potentially a class}'' -- G7). Some participants proposed inserting a slider at suggestion time to control the level of detail of the suggestions, allowing developers to switch from minimal to very long suggestions according to their needs (``\emph{Let the user choose what to suggest, for example with a slider}'' -- G2). 
    \item \textbf{Minimal suggestions with details asked on-demand} (4 groups): there was no single agreement on the minimum number of tokens that should be suggested; however, there was a certain agreement on preferring, initially, minimal suggestions with details asked on-demand to foster the user's reasoning (``\emph{Only suggest the function name to insert, so that {[}the developer{]} can solve the problem on their own {[}...{]} pieces of code with comments that allow us to receive cues/hints; do not suggest too much code.}'' \emph{--} G6).
    \item \textbf{Different alternatives} (3 groups): the aspect of having different alternatives offered by the code completion tool was also brought up as a preferable feature to improve precision in suggestions, particularly in the case of coding challenges that the developer has never faced (``\emph{Better to have many suggestions for new things, but for more `precise' things it's better to have single suggestions}'' -- G1).
    \item \textbf{Code accompanied by documentation} (2 groups): a  cross-cutting aspect of the above concerns the presence of documentation together with the code to improve comprehension and facilitate faster decision-making about the acceptance or rejection of the suggestion (``\emph{Include references to documentation for functions}'' -- G5).
    \item \textbf{Code adapted to the context} (3 groups): code should be aware of the context in which it is suggested , including variable names or coding styles relevant to the current project. According to the participants' proposals, the contexts could be represented by the developer's profiles (e.g., previous projects stored in repositories) or the projects already developed by the company (``\emph{{[}define{]} the standard of the suggested code, the style (in line with my style, or with the style of my company, of other repositories)...{}}'' -- G4).
\end{itemize}

\paragraph{How to customize suggestions}
The last aspect investigated during the study concerned how to customize the suggestions. Several suggestions emerged:
\begin{itemize}
    \item \textbf{Personalize activation timing of suggestions and explanations} (6 groups): the first need proposed by users concerns personalizing \textit{when} to suggest explantions to allow each developer to tailor the responsiveness of the tool to their own workflow (``\emph{The} \emph{activation timing, to disable/enable automatic hints}'' -- G5). 
    The need of customizing when to show explanations also emerged to adapt to different expertise levels of programmers (``\emph{Automatic generation of explanations is fine for novices, but could be obnoxious or even `outrageous' for experts}'' -- G2).
    \item \textbf{Customize appearance} (6 groups): another aspect that emerged in this dimension concerns customization of appearance: for example, many participants emphasized the need to customize the fonts, colors, and text size of the suggestions to ensure that they are visually distinct from the user's code according to their preferences (``\emph{Customize character size, color, font of suggestions..}'' -- G4).
    \item \textbf{Customize granularity of suggestions and explanations} (6 groups): in the What dimension, we reported the strategy of controlling granularity, which generally prescribes proposing the suggestion with different levels of granularity. Here emerged that users should be able to customize this option, e.g., the user could choose a 3, 4, 5 level scale of granularity, and for each of them, users can choose what should be suggested, e.g., the statement the user is writing, the control structure, the function, the class, or others (``\emph{Set the level and scope of suggestions (the context of action -- functions, loops, operators)}'' -- G7). 
    The need for customization also emerged for the granularity of explanations, to adapt to the user's knowledge by generating more or fewer comments to explain the suggested code (``\emph{... the presence of automatic comments and their quantity/frequency (every line, every function, etc.)}'' -- G5; ``\emph{Customize what is generated as explanations...{} Customization can be simple, such as a `what level of programming are you?' initial question}'' -- G6). 
    \item \textbf{Tailor coding style} (4 groups): participants stated that developers should adapt the tool to a certain coding style by fine-tuning it by feeding in, e.g., some of their own code repositories or their companies' past projects as examples (``\emph{Which context to use in the code generation as a `background training' ... set what repositories to consider}'' -- G2). Moreover, the developers should also be able to define quality standards to which the suggested code should adhere (``\emph{Control the quality of generated code, adherence to standards (even security ones)}'' -- G4).
\end{itemize}

\subsection{Explanations of the completion}\label{explanations-of-the-completion}

\begin{figure}[t]
    \centering
    \includegraphics[width=\linewidth]{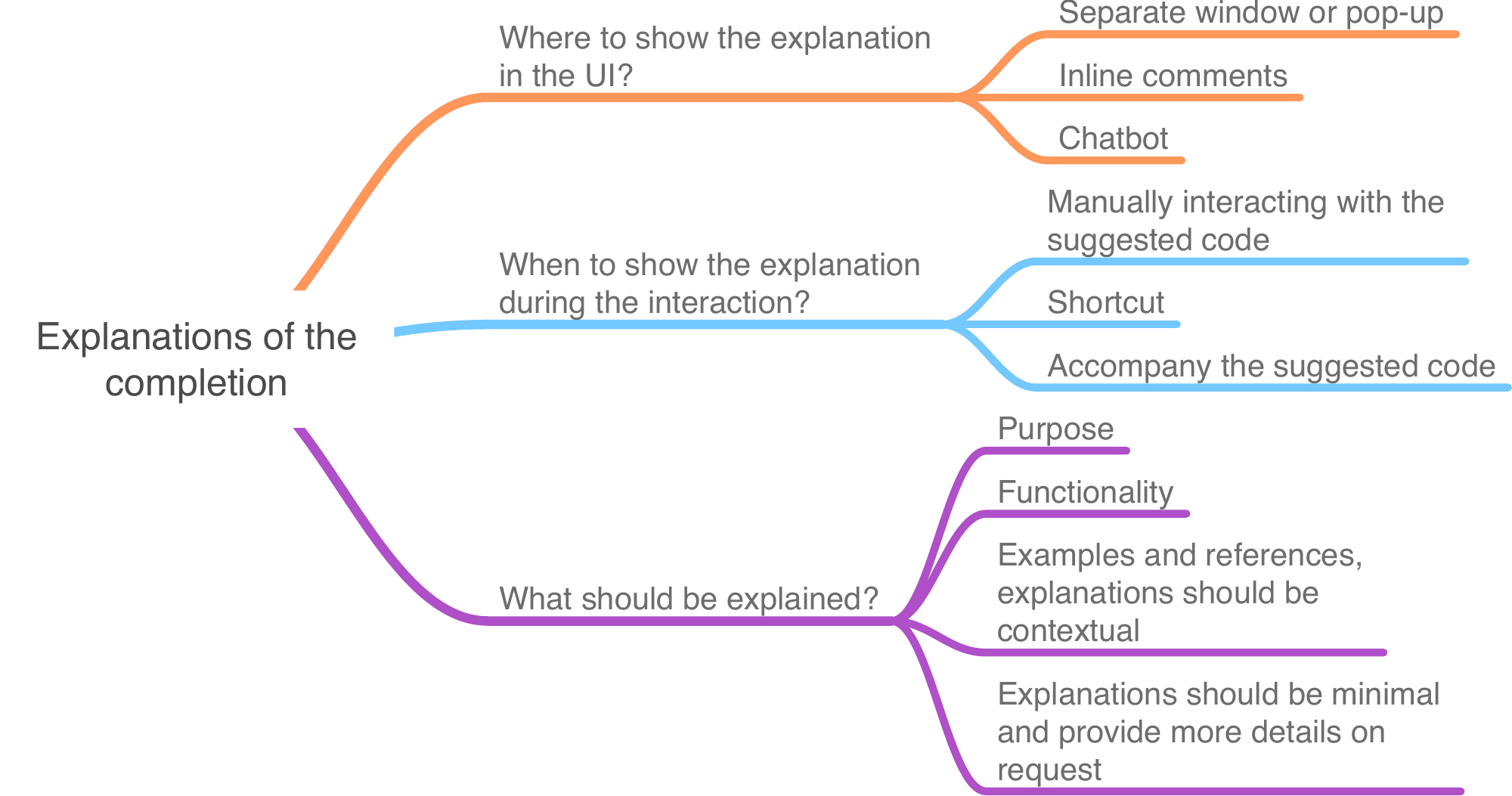}
    \caption{Themes emerged by analyzing the developers' answers to the questions on explanations produced by CCTs}
    \label{fig:findings-explanations}
\end{figure}

The results of this subsection answer \ref{RQ2}, which investigates developers’ expectations on how explanations should be delivered by CCTs. The study findings clarify user preferences about the timing, format, and content of explanations. In addition, the need for a customizable level of detail and contextual delivery answers \ref{RQ3}, highlighting the importance of tailoring explanations to user expertise and task demands. \Cref{fig:findings-explanations} represents the identified themes.

\paragraph{Where to show the explanation in the UI}
The first aspect we analyzed regarding suggested code explanations concerns \textbf{where} the explanations should be displayed in the user interface of the \acp{CCT}. Three suggestions emerged:

\begin{itemize}
    \item \textbf{Separate window or pop-up} (7 groups): several participants proposed that the explanations are better suited to a separate interface element, such as a sidebar or pop-up, allowing them to access the explanations without cluttering up the coding work area (``\emph{The} \emph{explanation and the function must be shown separately, otherwise there's too much text}'' -- G1).
    \item \textbf{Inline comments} (6 groups): another strategy concerns visualization employing inline comments within the code, thus providing an immediate context for the suggested code without having to navigate elsewhere (``\emph{As an expandible text under the code'' --} G7); however, this solution was indicated as viable only if the explanation length was limited, in order to avoid confusion in the code (``\emph{No comments, as they tend to be too long...{} but a `title' of the suggestion can be kept as a brief explanation in a comment}'' -- G6). 
    \item \textbf{Chatbots} (3 groups): some participants suggested accessing the explanations via a chatbot, allowing users to query the system in a conversational manner to obtain further clarification of the suggestions (\emph{``An alternative could be the chatbot to which I ask why {[}the code was suggested{]} after selecting the code''} -- G8).
\end{itemize}

\paragraph{When to show the explanation during the interaction}
Three complementary strategies emerged concerning the timing of explanations, i.e., when explanations should be shown during the interaction with the \acp{CCT}:
\begin{itemize}
    \item \textbf{Manually interacting with the suggested code} (6 groups): the first solution is to activate the explanations by manually interacting with the suggested code. For example, users can select the piece of suggested code they want to explain, right-click, then choose an ad-hoc function from a contextual menu, and the explanation is then displayed in a separate window, in the chatbot, or as a comment (``\emph{For example with the right click on the code and then pressing `explain'}'' -- G1); hovering over suggested code to show explanations contextually was also proposed by some groups (``\emph{... and then with an explicit request ...{} hovering on the code}'' -- G6).
    \item \textbf{Shortcut} (3 groups): an alternative solution concerns activation via shortcuts. Indeed, some participants suggested activating the explanations through a dedicated keyboard shortcut, granting users control without interfering with their workflow (``\emph{With a key I invoke the explanation}'' -- G8). 
    \item \textbf{Accompany the suggested code} (3 groups): an automatic approach along with the suggestion has also been proposed, i.e., also accompanying the suggested code with an automatic short explanation, expandable on-demand, which is especially useful when the suggestion is complex, reducing cognitive effort by presenting both simultaneously. A longer suggestion should be provided on-demand, for example, by interacting with the short suggestion and invoking more details (``\emph{... generated together with the code, as long as it is brief}'' -- G2; ``\emph{{[}explanation{]} details are provided on-demand}'' -- G3).
\end{itemize}

\paragraph{What should be explained}
The last aspect concerns what to explain. The desired content of the explanations was varied, with participants identifying three key elements: 

\begin{itemize}
    \item \textbf{Purpose} (5 groups): explain why a certain code suggestion was given. Indeed, many participants emphasized the need for explanations that clarify the purpose of the suggested code, specifying what it achieves in the broader context of the program (``\emph{{[}explain{]} why did it give me this suggestion?}'' -- G6; ``\emph{{[}explain{]} pros and cons of a suggested technique}'' -- G8).
    \item \textbf{Functionality} (4 groups): describe how the suggested code fulfills its purpose by reporting technical details such as the data structures, code constructs, data types and syntax involved (``\emph{What is the code composed of (in terms of data structures and chosen constructs)?}'' -- G6; ``\emph{explain the code, data types, ...{} syntax}'' -- G3); some groups also suggested displaying flow diagrams or pseudo-code to help developers understand complex suggestions and visualize the inner workings of the generated code (``\emph{With pseudocode, maybe also flowcharts}'' -- G1).
    \item \textbf{Examples and references} (5 groups): another popular aspect concerns the use of examples and references to sources : participants recommended that explanations include examples of the use of similar codes or references to reliable sources that might enable users to assess the credibility of the suggestion (``\emph{It may be sufficient to provide simple `pointers' to something that helps me understand ...{} such as links to source codes, or similar code examples as `insights'}'' -- G2).
    \item \textbf{Explanations should be contextual} (4 groups): regarding the scope of explanations, they should be contextual and not regard the entirety of the suggested code, but only the pieces of code that the user indicates (``\emph{We're not interested in being explained everything necessarily, but only of a piece {[}of code{]}}'' -- G1).
    \item \textbf{Explanations should be minimal and provide more details on request} (4 groups): regarding the length of explanations, there was a popular agreement that explanations should be minimal and provide more details on request. Participants suggested that this can be achieved by, e.g., clicking a button to expand the explanation preview or by interacting with a chatbot to obtain an explanation incrementally (``\emph{If something else specific is needed, the user can ask explicitly}'' -- G3).
\end{itemize}

\section{Discussion}\label{discussion}

\begin{figure*}[t]
    \centering
    \includegraphics[width=\linewidth]{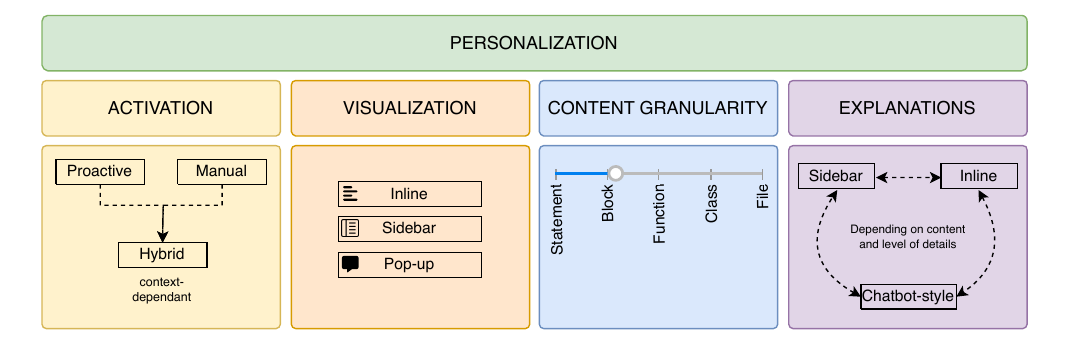}
    \caption{Graphical representation of users' preferences and mental model. The four aspects of \acp{CCT}, namely activation, visualization, content granularity, and explanations, are shown. All these dimensions are relevant for users' personalization.}
    \label{fig:mental-model}
\end{figure*}

Our findings highlight important points regarding developer preferences and requirements while working with \ac{AI}-driven \acp{CCT}. In particular, this work underlines the importance of designing \acp{CCT} for developers' mental models to promote productivity, efficiency, and an intuitive coding experience. \Cref{fig:mental-model} summarizes users' preferences and requirements, attempting to elicit developers' mental model of \acp{CCT}.

This section interprets and presents the findings of our study in light of the research questions. Specifically, \cref{ui-for-code-completion,explanations-of-the-completion} addressed \ref{RQ1}, \ref{RQ2}, and \ref{RQ3} by identifying how developers mentally model their interactions with AI-driven CCTs, what they expect from explanations, and how they express their need for personalization. More broadly, this section addresses \ref{RQ4} by distilling the study results into actionable human-centered design suggestions, which are intended to guide the development of CCTs that align with developers’ mental models and support diverse coding workflows.

\subsection{Mental model of activation and control (RQ1)}\label{activation-and-control}

Participants expressed different preferences regarding when code suggestions by \ac{CCT} should be triggered, namely proactive, manual, and hybrid activation. This aspect is supported by the results detailed in \cref{ui-for-code-completion} (`When to show suggestions?') and \cref{tab:table-rq2}, where proactive activation was favored by 4 groups, manual by 5 groups, and hybrid by 6 groups. 

This diversity in the mental model points to a key insight: in different contexts and coding tasks, developers might prefer different degrees of control. Some users mentioned that proactive suggestions, automatically activated during the coding flow breaks, might increase coding speed without explicit request. However, most users expressed the need for activation on-demand to control exactly when suggestions appeared. This is in line with similar studies regarding activation, as frequently \acp{CCT} are not adopted as they might disrupt developers' workflow \citep{Sergeyuk2025Using}. Rather, users favored a hybrid approach in which proactive behavior occurs only when a clear contextual need is present (such as when completing a complicated bit of code). 
However, proactive behavior should only occur once there is enough typing context (e.g., suggest code only if the file contains $X$ tokens or more) to avoid generating imprecise or random suggestions. 
Therefore, a dynamic and adaptive activation model could serve different user needs by altering its behavior in response to task complexity and user-characteristic usage patterns. Such a model would leave developers in control of their workflow while helping proactively when needed.

\subsection{Visualization preferences (RQ1)}\label{visualization-of-suggestions}

The visualization mechanism of suggestions impacts the usability and acceptance of \acp{CCT}. The participants proposed different visualization strategies, each with pros and cons. As shown in \cref{ui-for-code-completion} (`How to show suggestions?') and summarized in \cref{fig:findings-ui}, the majority of groups (5/8) preferred inline suggestions, as well as suggestions displayed in sidebars (7/8). Having inline suggestions can be useful to avoid losing the coding focus and allow the user to accept or reject the suggestion immediately. However, the inline solution is well-suited for short and concise suggestions that do not interfere too much with the code manually written by the users; on the contrary, in the case of long and complex suggestions, it may be better to have separate sidebars to limit visual clutter. A further level of control would be having users manually invoke popups containing the suggestion by clicking an icon shown close to the handwritten code (proposed by 2 groups). Moreover, a chatbot in a separate window could be used as an alternative way to suggest code on demand (proposed by 2 groups), with all the attached benefits of using conversational agents for code generation \citep{Ross2023Programmers}.

All these results may reveal an \emph{adaptable visualization}, where users can choose or alternate between inline, sidebar, and popups, and an \emph{adaptive visualization}, where the system visualizes the suggestion according to factors such as the suggestion length. Such modalities can improve usability by adapting to different workflows and developer preferences and might reduce the errors due to wrong intention communication as multiple choices may be provided when needed \citep{Sergeyuk2025Using}. This flexibility can also lower the cognitive overhead caused, for instance, by long pieces of generated code shown next to user-written code; in this way, \ac{CCT} allows developers to adjust the display of suggestions according to the complexity of the coding task or their personal coding style.

\subsection{Granularity and content of suggestions (RQ1, RQ3)}\label{content-and-granularity}

The study findings revealed that \acp{CCT} should be capable of generating suggestions at varying levels of granularity and that this granularity must be in the control of the users. This finding traces to \cref{ui-for-code-completion} and \cref{tab:table-rq4}, where it is reported that 7 out of 8 groups supported user control over suggestion granularity, including requests for minimal suggestions with additional detail on-demand found in 4 groups.

Participants did not clearly agree on the minimum or maximum number of tokens that should be suggested at any given time. Consequently, the \acp{CCT} should be designed to produce suggestions at different scopes, from predicting the next token in the code to completing control structures and generating entire functions or classes.

Although longer suggestions are useful, they can overwhelm the developer if displayed without considering the task context. For this reason, participants prefer retaining control over the level of granularity of suggestions. The necessity for varying degrees of granularity is also shared by other participants in different studies \citep{Wang2023How}. However, in general, users expressed a preference towards minimal, ``lower-scope'', suggestions at the first moment, with the possibility to obtain a longer, ``higher-scope'', suggestion on-demand; this approach of ``letting the user reason'' could help to prevent overreliance and deskilling. In practice, these directives might translate concretely into a minimalist suggestions behavior by default, with the possibility of using a slider or a control mechanism to adjust granularity in real time and adapt the suggestions to the contextual needs of the developer.

The control of granularity should also consider the number of offered alternatives; in certain use contexts, such as when developers face coding problems that they are not confident in, it may be beneficial for the user to have multiple options. Furthermore, code suggestions in such contexts may include documentation for constructs or functions the developer is not used to handling. Conversely, a suggestion with limited or no accompanying documentation should be presented for more straightforward and routinary tasks. These aspects could enhance user trust and overall acceptance of the suggestions.

\subsection{Preferences for explanation delivery (RQ2)}\label{explanation-preferences}

Participants generally considered explanations useful, but the answers to the question of \emph{what} should be contained in an explanation revealed a need for special care in their design. The concept of what defines a good explanation is not trivial and is an aspect that is widely studied in the literature \citep{Miller2019Explanation,Holzinger2020Measuring}. Moreover, even well-designed explanations should be evaluated with users to measure their quality \citep{Holzinger2020Measuring}. This elicitation study can be a precious source of information in understanding the desiderata of explanations in \acp{CCT} from the developers' point of view. A key finding is that explanations should be classified into two categories: \emph{why-explanations} and \emph{what-explanations}. Why-explanations address the rationale or purpose behind the generated code, whereas what-explanations concentrate on the functionality and technical intricacies of the code suggestion. The content should, however, be minimal, with the possibility to obtain more details when requested. This might translate into the \ac{CCT} by initially showing a brief why-explanation, providing a bird-eye view of what the generated code is for. A more thorough, on-demand explanation would instead detail the code nuances and technical aspects to help the developer get a deeper understanding of the suggestion, possibly even with flowcharts. Detailed explanations should also include references to trusted sources, such as documentation snippets, book pieces, etc., and even examples of use to enhance understanding.

Regarding where to show explanations, participants typically preferred separate elements (e.g., popups, sidebars), mainly because they did not want the explanations to be invasive and occupy space within the coding area. However, inline explanations may be accepted as comments close to the suggested code as long as their length is very limited. However, separate windows may allow developers to initiate an explanation process with a conversational agent, similar to what is possible to achieve with tools such as ChatGPT \citep{ChatGPT}.

Timing also plays a crucial role in showing the explanation. Some participants stated that explanations should be triggered manually using shortcuts, mouse clicks, or mouse hovering; notably, selecting code to explain with the mouse was a popular preference, also because it allows limiting the explanation to the context of the selection. Such an approach would have the advantage of reducing the amount of non-relevant explanation text to show to the user; moreover, it would also allow asking for further detailed explanations, limited to a specific piece of manually selected code, possibly to a chatbot. Some other participants instead expressed that brief explanations should be shown automatically (especially for novice developers), letting the user choose whether to receive more details on demand. Thus, all these findings indicate that adaptive, context-aware explanations can increase the understanding and trust of a tool dealing with complex or ambiguous code suggestions.

Participant preferences for delivering explanations are evidenced in \cref{explanations-of-the-completion}, \cref{tab:table-rq6} (7/8 groups favoring a separate window or pop-up), \cref{tab:table-rq7} (6/8 groups preferring manual interaction), and \cref{tab:table-rq8} (4/8 groups requesting minimal explanation with more details on request). 

\subsection{Personalization and design needs (RQ3, RQ4)}\label{personalization-needs}

An aspect that emerged as essential to fit the CCT's behavior in different mental models is the explicit personalization of the CCT. There was, indeed, little consensus among participants towards one universally accepted solution; in fact, many individual factors, such as experience with coding, quality requirements, and personal preferences, play a role in determining the optimal behavior (and, therefore, acceptance) of the CCT. For example, developers want to customize when suggestions should appear, how to visualize them, and which granularity should be used by default. Choosing whether suggestions should also be suggested automatically resulted in being, indeed, crucial to users for the acceptance of the CCT, while they also expressed the preference to edit what manual action (e.g., shortcut, toggle button, etc.) activates the code completion. In addition, customizing the suggestions' font, colors, and size and controlling the suggestions' position within the \ac{IDE} also proved useful. Another critical point regarded the need to customize the granularity of suggestions to control the amount of code provided by the tool, both regarding the scope of the suggestion (function-level, control structure-level, etc.) and the number of provided alternatives.

The users expressed the need to customize aspects of generating explanations regarding the suggested code. Therefore, developers must be able to customize aspects such as the activation timing of explanations (whether to provide them proactively) and their granularity (such as the amount and frequency of comments within the generated code and the explanation abstraction).

Finally, customization of a \ac{CCT} should include the choice of default stylistic code (such as coding conventions, variable names, and styles) when generating code suggestions, possibly adapting to other past projects. Therefore, the developer should be allowed to ``fine-tune'' the code-generating model with a selection of other code repositories to align the suggestions with a knowledge base representing the generated code's style. This customization should also allow the developer to set standards of quality and security to which the generated code has to comply.

Thus, a personalization layer in \acp{CCT} should be introduced to allow developers to define \ac{CCT} settings to improve the tool's usability, acceptance, and trust. These customization options may aid in implementing a smoother running and adaptable \ac{CCT} experience to strengthen the correspondence of the \ac{CCT} conduct with developers' mental models.

\section{Reifying the mental model: the ATHENA prototype}\label{prototype-development}

In this section, we put the mental model derived from our elicitation study into practice by presenting ATHENA, a prototype implementation that embodies and operationalizes the key design principles and the identified user interaction preferences.

ATHENA (AI Tool for Human-centered ENhanced coding Assistance) is a proof-of-concept \ac{CCT} for Visual Studio Code \ac{IDE}, developed to verify and demonstrate the feasibility of the actionable guidelines emerging from our study. The implementation incorporates high configurability, such as suggestion timing, granularity control, adaptive visualization (inline, sidebar, popup), and customizable explanation delivery that allows developers to tailor the extension to their preferences and workflows. As its namesake, Athena, the Greek goddess of skill and strategy, the tool aims to provide developers with precise, adaptable coding support.

The extension dynamically adapts to the user's coding preferences and environment, ensuring seamless integration into various workflows. The AI backend employs OpenAI's \textit{GPT-4o mini}\footnote{https://openai.com/index/gpt-4o-mini-advancing-cost-efficient-intelligence/} to generate code suggestions and explanations. Of course, the AI model can be replaced with other similar solutions.

One of the major takeaways from the user study is that no ``one-size-fits-all'' approach exists to accommodate the needs of different types of users. Therefore, high customization was designated as a core feature in the tool's design. To this aim, \textit{ATHENA} offers robust configurability through its \textit{Settings Wizard Panel}, which is automatically displayed upon installation and remains accessible via a customizable shortcut. Settings are managed through Visual Studio Code's built-in infrastructure, supporting workspace and user-level configurations. The wizard is divided into two main sections: \textit{General Settings} and \textit{Shortcut Configuration}.

\begin{figure}[t]
    \centering
    \includegraphics[width=.75\linewidth]{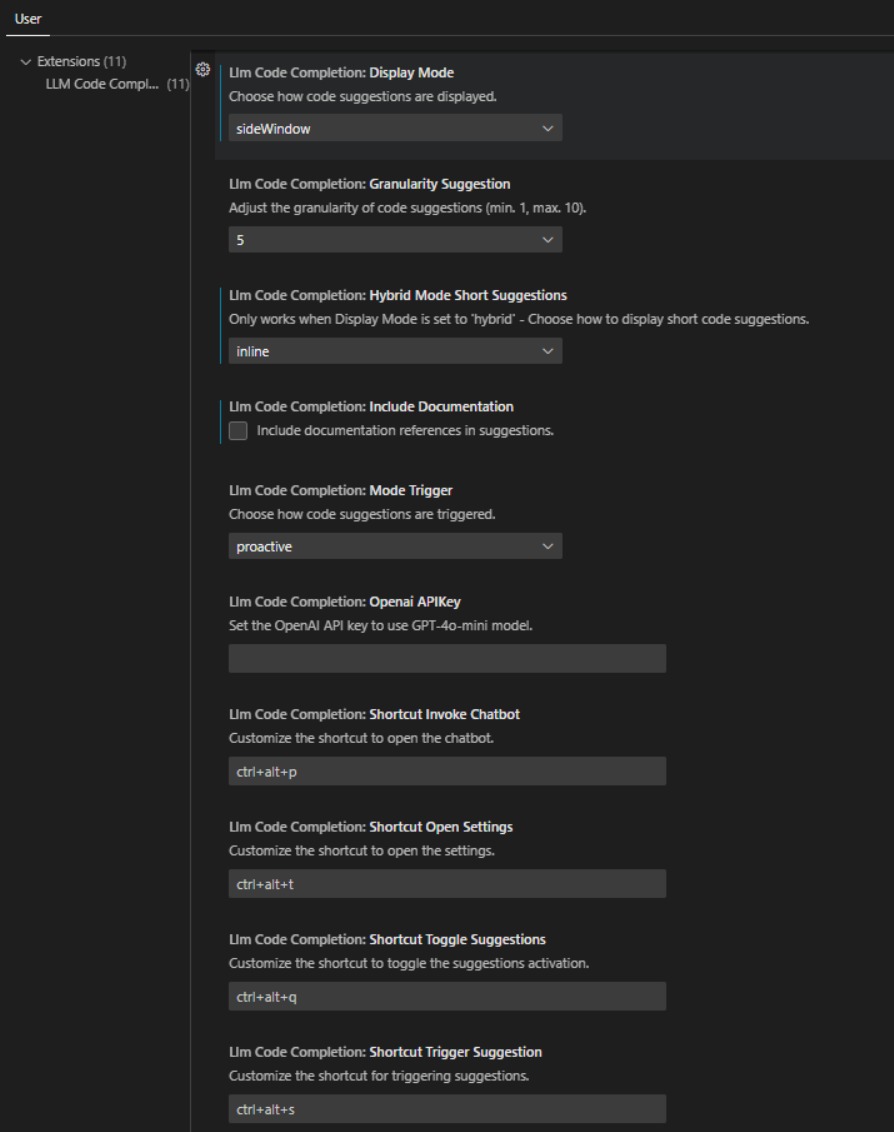}
    \caption{Screenshot of the settings provided with the prototype.}
    \label{fig:settings}
\end{figure}

The general settings (Figure \ref{fig:settings}) enable users to customize the behavior and appearance of the CCT:

\begin{itemize}
    \item \textit{Suggestion Length}: A drop-down menu with options ranging from 1 to 10 adjusts the granularity of suggestions, balancing concise completions with detailed recommendations tailored to the user’s needs.    
    \item \textit{Suggestion Appearance}: Users can choose from four display modes:
    \begin{itemize}
        \item Inline: Embeds suggestions directly in the text editor for a seamless experience.
        \item Tooltip: Shows suggestions in a hover menu near the cursor.
        \item Side Window: Displays suggestions in a dedicated subpanel for enhanced focus.
        \item Hybrid: Combines inline/tooltip suggestions for short completions and the side window for longer or more complex suggestions. 
    \end{itemize}
   \item \textit{Suggestion Timing}: Users can toggle between:
   \begin{itemize}
       \item Proactive: Automatically triggers suggestions as the user types, ideal for uninterrupted workflows.
        \item Manual: Requires explicit activation through a shortcut, reducing distractions.
   \end{itemize}
   \item \textit{Source References}: A checkbox lets users include or omit sources of suggestions, ensuring transparency for users who value understanding the origin of generated code
   \item \textit{Comments Frequency}: A slider allows users to adjust the frequency of comments in generated code, catering to both minimalist and documentation-focused coding styles.
\end{itemize}



The shortcut configuration section, on the other side, allows users to assign shortcuts to key actions:
\begin{itemize}
    \item \textit{Invoke chatbot}: Quickly access a persistent chatbot for coding queries, explanations, and guidance.
    \item \textit{Open settings panel}: Revisit the settings at any time to adjust preferences.
    \item \textit{Trigger manual suggestions}: Activate suggestions when working in manual mode.
    \item \textit{Toggle auto-completion}: Enable or disable automated suggestions dynamically.
\end{itemize}

After the system is installed and customized, it can be used within Visual Studio Code. To assist developers during coding activities, ATHENA provides different interaction modalities. The first one is completely automatic and consists of proactive suggestions, which are automatically produced by ATHENA after a short idle time, and according to the preferences expressed by the users. If the developer wants to be more in control of the suggestions, ATHENA provides two different solutions. In the first case, shortcuts can be triggered while writing pieces of code, while in the second case, a chatbot can be open to interactively receive suggestions. 

ATHENA allows the developer to receive explanations for selected pieces of code within the editor via contextual IDE options. Specifically, a light-bulb icon appears at the top of the selected code, allowing for three distinct actions:
\begin{itemize}
    \item \textit{Explain the Code (Concise)}: Offers a concise summary of the code's functionality for the selected code. 
    \item \textit{Explain the Code (Detailed)}: Offers a deeper explanation of the purpose and structure of the selected code. This can also include the use of examples.
    \item \textit{Open Chatbot}: Provides direct access to the chatbot with the selected code as context for further assistance.
\end{itemize}

\begin{figure}[t]
    \centering
    \includegraphics[width=.8\linewidth]{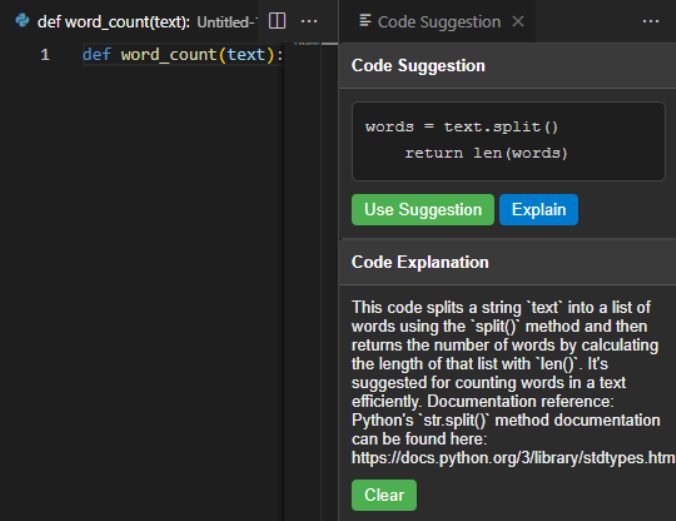}
    \caption{ATHENA interface: the user prefers to show suggestions in a separate panel.}
    \label{fig:sidebar}
\end{figure}

Regardless of the solutions used by the developer to invoke the code suggestions and explanations, ATHENA's output can be visualized in two ways, according to the developer's preferences. The first solution consists of showing code/explanations within the code editor (either inline or in a tooltip). Another option is to show suggestions and explanations separately from the code editor within a side panel, as shown in Figure \ref{fig:sidebar}.

\begin{itemize}
    \item \textit{Suggestions/Completions}: Displays code completions only when the user has configured the display mode to ``Side Window'' or ``Hybrid.''
    \item \textit{Explanations}: Provides detailed explanations for selected code, ensuring clarity and context for the developer's work.
\end{itemize}

Given the high degree of customizability, \textit{ATHENA} addresses the issue of cold start by prompting the user with a question regarding their programming experience upon initial execution. Based on the user's response (which may be ``beginner'', ``intermediate'', or ``advanced''), the tool adapts the default values in order to tailor its behavior. For instance, the results of the elicitation study suggest that activation should be on-demand for intermediate users, as this was the most preferred behavior overall. On the other hand, a novice user might benefit from a proactive approach that suggests code as they type. As another example, an intermediate user might benefit from the inclusion of documentation references in the suggestions, whereas an expert user might perceive them as superfluous or cumbersome. These default settings configurations were chosen arbitrarily; however, future work may inform the design of the optimal configurations for different categories of users. Nevertheless, after the initial question, the user is presented with the settings panel, where each configuration can be readily modified. 

\subsection{Use case scenarios}
In the following, two scenarios are reported to clarify how the ATHENA tool can be installed and used the first time (Scenario 1, \Cref{scenario-1}) and how the tool can be used by the same users in different contexts, thus demonstrating its suitability to different mental models of the same users (Scenario 2, \Cref{scenario-2}). Two videos of the two scenarios are also attached to show how ATHENA works.

\subsubsection{Scenario 1: ATHENA installation and first usage}\label{scenario-1}

Alex, a junior software developer at CodeCraft, is starting a new back-end API project for an e-commerce platform. To streamline their workflow, Alex installs the ATHENA Extension in Visual Studio Code.
After installation, the extension prompts Alex with a popup asking their level of experience with programming. Alex answers the question by indicating ``intermediate'': this will automatically set the tool's parameters to an intermediate user profile, e.g., by disabling proactive suggestions. 
Nonetheless, the tool launches its \textit{Settings Panel}, guiding Alex through the initial configuration process and giving them the possibility to change the default parameters. In the general settings, Alex sets the display mode to \textit{Inline} to benefit from suggestions directly placed inside the editor. Alex thinks that suggestions that appear automatically while coding would be better and, therefore, chooses to set the suggestion trigger mode to \textit{Proactive}. To fine-tune the experience, Alex adjusts the \textit{suggestion length} slider to a medium level for balanced granularity.
In the \textit{Shortcut Configuration} section, Alex assigns shortcuts for invoking the chatbot, toggling auto-completion, and manually triggering suggestions. Satisfied with the setup, Alex saves the preferences and begins coding.
As Alex starts writing the first few lines, inline suggestions appear seamlessly, offering useful completions for common coding patterns. 

\subsubsection{Scenario 2: adapting to debugging needs}\label{scenario-2}

A few weeks later, Alex is tasked with debugging a critical issue in the API's payment processing module. While the \textit{Proactive Mode} had been helpful during development, Alex finds it distracting when troubleshooting errors. Recognizing the extension's flexibility, Alex opens the Settings Panel and switches the trigger mode to \textit{Manual}, ensuring suggestions only appear when explicitly requested. Alex also reduces the \textit{Suggestion Length} to receive concise completions that are better suited for debugging.

During debugging, Alex selects a block of code that seems problematic. A lightbulb icon appears next to the selected code, offering three options: ``Explain the Code (Concise)'', ``Explain the Code (Detailed)'', and ``Open Chatbot''. Alex selects the first option (\textit{Explain the Code (Concise)}) and promptly receives a brief summary of the code's functionality. Alex wants to delve deeper into the details of the generated code; therefore, Alex selects the second option (\textit{Explain the Code (Detailed)}), which provides a more in-depth explanation, to help the people understand the logic and potential pitfalls. To receive further guidance, Alex selects the \textit{Open Chatbot} option; this invokes the chatbot, which provides context-aware suggestions for refactoring the code. 

Later, Alex notices the Side Panel's explanation subpanel automatically updating with insights about the selected code. Combined with the concise suggestions displayed in the completion subpanel, Alex efficiently resolves the issue.

By tailoring the tool’s behavior to their specific debugging needs, Alex maintains the focus and leverages the extension's features to complete the task effectively. The combination of personalized settings, code actions, and a context-aware chatbot reinforces the tool's versatility and value in diverse workflows.

\section{Threats to validity}
\label{threats}

Even if our study provides useful insights about developers' mental models on CCTs, several limitations must be acknowledged.

\paragraph{External validity} The participants involved in our study were recruited among the students of two Italian universities, primarily third-year undergraduate computer science students. Although this sampling is appropriate for our exploratory goals, it does not represent the broader population of professional developers. Moreover, the sample skewed heavily male (48 out of 56), and few participants had extensive industry experience. In addition, students may lack the experience necessary to generate highly innovative or realistic design expectations. As a consequence, the generalizability of the findings to different developer populations is limited. Future research is needed to validate and extend these insights through studies involving participants from different industry settings and different skills, with more females, coming from multiple countries, and with more varied demographic backgrounds.

\paragraph{Internal validity} The workshop methodology adopted in our study, while useful for eliciting user mental models, can introduce biases. For instance, some participants may dominate the discussion while others may be quieter and less participatory. To mitigate this bias, we structured the workshops introducing an individual reflection phase for each question, and encouraged all participants to voice their opinions before group synthesis. Another potential problem regards the fact that researcher facilitation may still have unintentionally shaped some discussions. To mitigate this bias, audio recordings have been triangulated with observer transcription before performing the qualitative analysis.

\paragraph{Construct validity} The eight questions used in the workshops to guide the discussion were derived from the analysis of the existing literature on CCTs, with the aim of covering the critical interaction dimensions of CCTs. This structure supported a systematic coverage of relevant aspects; however, it may have constrained participants from reporting issues outside the predefined scope. To mitigate this problem, we introduced open-ended discussion and sketching tools that helped surface unanticipated questions.

\paragraph{Methodological limitations} Co-design workshops tend to foster breadth and richness of group discussion. However, participants, in particular students, may find it difficult to anticipate how they would interact with a real CCT, thus introducing a form of hypothetical bias. This is particularly critical for early-stage technologies, as in the case of our study. Furthermore, participants’ comments may not fully reflect their behavior in the real world. As future work, to mitigate this problem, complementary methods should be adopted, for example, individual interviews, in-the-wild deployment, and usage logging.

\section{Conclusion}\label{conclusion}

The use of \ac{CCT}s has seen a notable increase in recent years and chances are that they will be used more and more in the future. Nevertheless, their acceptance strongly depends on their alignment with developers' mental models. This study investigated the desiderata of \ac{CCT}s through an elicitation study with programmers of varying expertise levels. Our findings indicate that, except for a few shared properties, there is no consensus on features such as activation timing, position, and granularity of suggestions and explanations. This lack of agreement on the specifics of CCTs thus indicates that they necessitate a high degree of customizability.

Based on the findings of the elicitation study, we have also developed \textit{ATHENA}, a prototype of a \ac{CCT} for Visual Studio Code that is highly customizable and implements most of the features identified by our users. This tool is openly available to the research community and will be essential to conduct future user studies that assess how different characteristics of a \ac{CCT} (e.g., the position of the code suggestion or the activation behavior) can affect its usability and user performance.
In future work, we will investigate these differences with a controlled experiment to isolate different variables and measure their effects individually.

Limitations of this study mostly regard the convenience sampling methodology that was adopted, which led to a sample of very young participants, with limited to no expertise in companies. Therefore, despite the large sample size, the findings of this study are not directly generalizable to professional developers. To address this limitation we plan to conduct usablity studies with different categories of users. Such studies will be initially conducted in a laboratory setting, and then in-the-wild within an IT company; this will allow us to measure the effectiveness and usability of \textit{ATHENA} compared to other popular \ac{CCT}s. Moreover, we will observe how different features of a CCT can affect the usability and performance of users with varying degrees of expertise with different categories of users. 
Future work will also entail further developing our \ac{CCT} and extending its functionalities by implementing the remaining features that emerged from the elicitation study. Moreover, we will explore the use of different \ac{LLM}s as the AI code completion engine of the tool; in fact, to generate code suggestions and explanations \textit{ATHENA} employs OpenAI's GPT-4o mini, an \ac{LLM} which has a limited size but performs well on coding problems \citep{2024GPT4oMini}. However, different models may prove more suitable for use in corporate settings. For instance, a non-commercial solution such as Meta's open-source LLM \textit{LLama} \citep{2024LLama3-2} could be deployed on a company's server to prevent sharing confidential data, such as project repositories, with third parties. Furthermore, models of varying sizes could be contemplated in light of a tradeoff between costs and performance.

\section*{Acknowledgements}\label{sec:acknowledgements}
This research is partially funded by the Italian Ministry of University and Research (MUR) and by the European Union - NextGenerationEU, Mission 4, Component 2, Investment 1.1, under grant PRIN 2022 ``DevProDev: Profiling Software Developers for Developer-Centered Recommender Systems'' — CUP: H53D23003620006.

The research of Andrea Esposito is funded by a Ph.D.~fellowship within the framework of the Italian ``D.M.~n.~352, April 9, 2022'' - under the National Recovery and Resilience Plan, Mission 4, Component 2, Investment 3.3 - Ph.D. Project ``Human-Centered Artificial Intelligence (HCAI) techniques for supporting end users interacting with AI systems'', co-supported by ``Eusoft S.r.l.'' (CUP H91I22000410007).

The research of Francesco Greco is funded by a PhD fellowship within the framework of the Italian “D.M. n. 352, April 9, 2022”- under the National Recovery and Resilience Plan, Mission 4, Component 2, Investment 3.3 - PhD Project “Investigating XAI techniques to help user defend from phishing attacks”, co-supported by “Auriga S.p.A.” (CUP H91I22000410007).

\section*{Statements and Declaration}\label{statements-and-declaration}

\subsection*{Conflicts of Interest}\label{conflicts-of-interest}

The authors declare no conflicts of interest.

\subsection*{CRediT authorship contribution statement}
\credit{Giuseppe Desolda}{Conceptualization, Formal analysis, Funding acquisition, Investigation, Methodology, Project administration, Supervision, Validation, Writing -- original draft, Writing -- review \& editing}
\credit{Andrea Esposito}{Investigation, Methodology,  Validation, Visualization, Writing -- original draft, Writing -- review \& editing}
\credit{Francesco Greco}{Data curation, Formal analysis, Investigation, Methodology, Software, Writing -- original draft, Writing -- review \& editing}
\credit{Cesare Tucci}{Data curation, Formal analysis, Investigation, Methodology, Software, Writing -- original draft, Writing -- review \& editing}
\credit{Paolo Buono}{Writing -- original draft, Writing -- review \& editing}
\credit{Antonio Piccinno}{Writing -- original draft, Writing -- review \& editing}

\insertcreditsstatement

 \bibliographystyle{elsarticle-num-names}
 \bibliography{main}

\appendix
\section{Themes and codes}\label{app:themes}
\input{tables}

\end{document}

%% file: tables.tex
\begin{table}[H]
\centering
\caption{Themes for Question 1}
\label{tab:table-rq1}
\resizebox{\textwidth}{!}{%
\begin{tabular}{|cclc|}
\hline
\multicolumn{4}{|c|}{\textit{\textbf{Q1. Where to show the code completion in the UI?}}} \\ \hline
\multicolumn{1}{|c|}{\textit{\textbf{Theme}}} & \multicolumn{1}{c|}{\textit{\textbf{Theme frequency}}} & \multicolumn{1}{c|}{\textit{\textbf{Codes}}} & \textit{\textbf{Code frequency}} \\ \hline
\multicolumn{1}{|c|}{\multirow{3}{*}{\textbf{Separate window}}} & \multicolumn{1}{c|}{\multirow{3}{*}{\textbf{6}}} & \multicolumn{1}{p{7cm}|}{Code 1: "Generated code should appear in a side window or tab" sideWindow: 1,3,4,5} & 4 \\ \cline{3-4} 
\multicolumn{1}{|c|}{} & \multicolumn{1}{c|}{} & \multicolumn{1}{p{7cm}|}{Code 2: "Generated code should appear in a tab at the bottom of the IDE" bottomWindow: 3} & 2 \\ \cline{3-4} 
\multicolumn{1}{|c|}{} & \multicolumn{1}{c|}{} & \multicolumn{1}{p{7cm}|}{Code 3: "Generated code should appear in a side movable window" movableTab: 7} & 2 \\ \hline
\multicolumn{1}{|c|}{\multirow{2}{*}{\textbf{Popup}}} & \multicolumn{1}{c|}{\multirow{2}{*}{\textbf{3}}} & \multicolumn{1}{p{7cm}|}{Code 1: "Pop-up near the main code"} & 2 \\ \cline{3-4} 
\multicolumn{1}{|c|}{} & \multicolumn{1}{c|}{} & \multicolumn{1}{p{7cm}|}{Code 2: "Toggable pop-ups to view code completions"} & 1 \\ \hline
\multicolumn{1}{|c|}{\multirow{3}{*}{\textbf{Inline}}} & \multicolumn{1}{c|}{\multirow{3}{*}{\textbf{5}}} & \multicolumn{1}{p{7cm}|}{Code 1: Inline code shown as a preview} & 2 \\ \cline{3-4} 
\multicolumn{1}{|c|}{} & \multicolumn{1}{c|}{} & \multicolumn{1}{p{7cm}|}{Code 2: Inline suggestion with a different font to differentiate it} & 1 \\ \cline{3-4} 
\multicolumn{1}{|c|}{} & \multicolumn{1}{c|}{} & \multicolumn{1}{p{7cm}|}{Code 3: Completion code should appear inline but be explictly accepted by the user} & 2 \\ \hline
\multicolumn{1}{|c|}{\textbf{Adaptive visualization}} & \multicolumn{1}{c|}{\textbf{2}} & \multicolumn{1}{p{7cm}|}{Code 1: "Separate area for complex hints, inline for short ones"} & 2 \\ \hline
\end{tabular}%
}
\end{table}

\begin{table}[H]
\centering
\caption{Themes for Question 2}
\label{tab:table-rq2}
\resizebox{\textwidth}{!}{%
\begin{tabular}{|cclc|}
\hline
\multicolumn{4}{|c|}{\textit{\textbf{Q2. When should the suggestion be activated during the interaction?}}} \\ \hline
\multicolumn{1}{|c|}{\textit{\textbf{Theme}}} & \multicolumn{1}{c|}{\textit{\textbf{Theme frequency}}} & \multicolumn{1}{c|}{\textit{\textbf{Codes}}} & \textit{\textbf{Code frequency}} \\ \hline
\multicolumn{1}{|c|}{\multirow{2}{*}{\textbf{Proactive}}} & \multicolumn{1}{c|}{\multirow{2}{*}{\textbf{4}}} & \multicolumn{1}{p{7cm}|}{Code 1: "Trigger hint during typing automatically"} & 2 \\ \cline{3-4} 
\multicolumn{1}{|c|}{} & \multicolumn{1}{c|}{} & \multicolumn{1}{p{7cm}|}{Code 2: "Hints appear when user is blocked"} & 4 \\ \hline
\multicolumn{1}{|c|}{\multirow{5}{*}{\textbf{Manual}}} & \multicolumn{1}{c|}{\multirow{5}{*}{\textbf{5}}} & \multicolumn{1}{p{7cm}|}{Code 1: "Manual trigger with shortcut"} & 2 \\ \cline{3-4} 
\multicolumn{1}{|c|}{} & \multicolumn{1}{c|}{} & \multicolumn{1}{p{7cm}|}{Code 2: "Manual activation preferred over proactive activation"} & 2 \\ \cline{3-4} 
\multicolumn{1}{|c|}{} & \multicolumn{1}{c|}{} & \multicolumn{1}{p{7cm}|}{Code 3: "Manual activation by selecting code"} & 2 \\ \cline{3-4} 
\multicolumn{1}{|c|}{} & \multicolumn{1}{c|}{} & \multicolumn{1}{p{7cm}|}{Code 4: "Manual activation option required"} & 3 \\ \cline{3-4} 
\multicolumn{1}{|c|}{} & \multicolumn{1}{c|}{} & \multicolumn{1}{p{7cm}|}{Code 5: "On-demand activation via comment"} & 1 \\ \hline
\multicolumn{1}{|c|}{\multirow{3}{*}{\textbf{Hybrid}}} & \multicolumn{1}{c|}{\multirow{3}{*}{\textbf{6}}} & \multicolumn{1}{p{7cm}|}{Code 1: "Automatic hints during typing, manual by selection"} & 5 \\ \cline{3-4} 
\multicolumn{1}{|c|}{} & \multicolumn{1}{c|}{} & \multicolumn{1}{p{7cm}|}{Code 2: "Hybrid: automatic for obvious hints, manual otherwise"} & 1 \\ \cline{3-4} 
\multicolumn{1}{|c|}{} & \multicolumn{1}{c|}{} & \multicolumn{1}{p{7cm}|}{Code 3: "Manual hints for user when blocked"} & 1 \\ \hline
\multicolumn{1}{|c|}{\textbf{When coding context is available}} & \multicolumn{1}{c|}{\textbf{4}} & \multicolumn{1}{p{7cm}|}{Code 1: "Hints appear after analyzing typing context"} & 4 \\ \hline
\end{tabular}%
}
\end{table}

\begin{table}[H]
\centering
\caption{Themes for Question 3}
\label{tab:table-rq3}
\resizebox{\textwidth}{!}{%
\begin{tabular}{|cclc|}
\hline
\multicolumn{4}{|c|}{\textit{\textbf{Q3. How should the suggested code be shown?}}} \\ \hline
\multicolumn{1}{|c|}{\textit{\textbf{Theme}}} & \multicolumn{1}{c|}{\textit{\textbf{Theme frequency}}} & \multicolumn{1}{c|}{\textit{\textbf{Codes}}} & \textit{\textbf{Code frequency}} \\ \hline
\multicolumn{1}{|c|}{\multirow{5}{*}{\textbf{Inline}}} & \multicolumn{1}{c|}{\multirow{5}{*}{\textbf{5}}} & \multicolumn{1}{p{7cm}|}{Code 1: "Change font to transparent for generated code"} & 3 \\ \cline{3-4} 
\multicolumn{1}{|c|}{} & \multicolumn{1}{c|}{} & \multicolumn{1}{p{7cm}|}{Code 2: Shorter suggested code is transparent} & 2 \\ \cline{3-4} 
\multicolumn{1}{|c|}{} & \multicolumn{1}{c|}{} & \multicolumn{1}{p{7cm}|}{Code 3: Inline suggestions for short code} & 2 \\ \cline{3-4} 
\multicolumn{1}{|c|}{} & \multicolumn{1}{c|}{} & \multicolumn{1}{p{7cm}|}{Code 4: "Color-coded suggestion types"} & 1 \\ \cline{3-4} 
\multicolumn{1}{|c|}{} & \multicolumn{1}{c|}{} & \multicolumn{1}{p{7cm}|}{Code 5: "Toggle color-coding for AI/human code"} & 1 \\ \hline
\multicolumn{1}{|c|}{\multirow{3}{*}{\textbf{Separate area}}} & \multicolumn{1}{c|}{\multirow{3}{*}{\textbf{7}}} & \multicolumn{1}{p{7cm}|}{Code 1: "Separate window for detailed suggestions"} & 6 \\ \cline{3-4} 
\multicolumn{1}{|c|}{} & \multicolumn{1}{c|}{} & \multicolumn{1}{p{7cm}|}{Code 2: "Separate place for code suggestions"} & 1 \\ \cline{3-4} 
\multicolumn{1}{|c|}{} & \multicolumn{1}{c|}{} & \multicolumn{1}{p{7cm}|}{Code 3: Suggestion must be not invasive} & 2 \\ \hline
\multicolumn{1}{|c|}{\textbf{Clicking an icon or badge}} & \multicolumn{1}{c|}{\textbf{2}} & \multicolumn{1}{p{7cm}|}{Code 1: "Pressable icon to signal suggestions availability"} & 2 \\ \hline
\multicolumn{1}{|c|}{\multirow{2}{*}{\textbf{Chatbot}}} & \multicolumn{1}{c|}{\multirow{2}{*}{\textbf{3}}} & \multicolumn{1}{p{7cm}|}{Code 1: "The code suggestion should happen as a dialogue with a chatbot"} & 2 \\ \cline{3-4} 
\multicolumn{1}{|c|}{} & \multicolumn{1}{c|}{} & \multicolumn{1}{p{7cm}|}{Code 2: Chatbot for detailed requests} & 1 \\ \hline
\end{tabular}%
}
\end{table}

\begin{table}[H]
\centering
\caption{Themes for Question 4}
\label{tab:table-rq4}
\resizebox{\textwidth}{!}{%
\begin{tabular}{|cclc|}
\hline
\multicolumn{4}{|c|}{\textit{\textbf{Q4. What should be suggested?}}} \\ \hline
\multicolumn{1}{|c|}{\textit{\textbf{Theme}}} & \multicolumn{1}{c|}{\textit{\textbf{Theme frequency}}} & \multicolumn{1}{c|}{\textit{\textbf{Codes}}} & \textit{\textbf{Code frequency}} \\ \hline
\multicolumn{1}{|c|}{\multirow{6}{*}{\textbf{Controlling granularity}}} & \multicolumn{1}{c|}{\multirow{6}{*}{\textbf{7}}} & \multicolumn{1}{p{7cm}|}{Code 1: "Code should be given at varying granularity, choosable by the user"} & 2 \\ \cline{3-4} 
\multicolumn{1}{|c|}{} & \multicolumn{1}{c|}{} & \multicolumn{1}{p{7cm}|}{Code 2: "Code should be generated at a scope level defined by the user"} & 4 \\ \cline{3-4} 
\multicolumn{1}{|c|}{} & \multicolumn{1}{c|}{} & \multicolumn{1}{p{7cm}|}{Code 3: "Generated code should be limited to the function scope  level"} & 1 \\ \cline{3-4} 
\multicolumn{1}{|c|}{} & \multicolumn{1}{c|}{} & \multicolumn{1}{p{7cm}|}{Code 4: "Generated code should be limited to the file scope  level"} & 2 \\ \cline{3-4} 
\multicolumn{1}{|c|}{} & \multicolumn{1}{c|}{} & \multicolumn{1}{p{7cm}|}{Code 5: "Generated code should have a minimum amount of tokens"} & 1 \\ \cline{3-4} 
\multicolumn{1}{|c|}{} & \multicolumn{1}{c|}{} & \multicolumn{1}{p{7cm}|}{Code 6: "Generated code should at least be one line of code"} & 1 \\ \hline
\multicolumn{1}{|c|}{\multirow{2}{*}{\textbf{Different alternatives offered}}} & \multicolumn{1}{c|}{\multirow{2}{*}{\textbf{3}}} & \multicolumn{1}{p{7cm}|}{Code 1: "Multiple code alternatives to choose from are preferred"} & 2 \\ \cline{3-4} 
\multicolumn{1}{|c|}{} & \multicolumn{1}{c|}{} & \multicolumn{1}{p{7cm}|}{Code 2: "In certain contexts, users should be able to choose from multiple code alternatives"} & 2 \\ \hline
\multicolumn{1}{|c|}{\multirow{4}{*}{\textbf{\begin{tabular}[c]{@{}c@{}}Minimal suggestions,\\ details on-demand\end{tabular}}}} & \multicolumn{1}{c|}{\multirow{4}{*}{\textbf{4}}} & \multicolumn{1}{p{7cm}|}{Code 1: "Code should not be given right away, but be shown as an expandable preview"} & 1 \\ \cline{3-4} 
\multicolumn{1}{|c|}{} & \multicolumn{1}{c|}{} & \multicolumn{1}{p{7cm}|}{Code 2: "Suggested code should be given incrementally to foster the user's reasoning"} & 3 \\ \cline{3-4} 
\multicolumn{1}{|c|}{} & \multicolumn{1}{c|}{} & \multicolumn{1}{p{7cm}|}{Code 3: "Single suggestions (with no alternatives) are preferred"} & 1 \\ \cline{3-4} 
\multicolumn{1}{|c|}{} & \multicolumn{1}{c|}{} & \multicolumn{1}{p{7cm}|}{Code 4: "The code completion could be limited to just one token"} & 1 \\ \hline
\multicolumn{1}{|c|}{\textbf{Codes and documentation}} & \multicolumn{1}{c|}{\textbf{2}} & \multicolumn{1}{p{7cm}|}{Code 1: "Show documentation together with the suggested code"} & 2 \\ \hline
\multicolumn{1}{|c|}{\textbf{Code adapted to the context}} & \multicolumn{1}{c|}{\textbf{3}} & \multicolumn{1}{p{7cm}|}{Code 1: "The generated code should adapt to the context of the user (user profile and their work environment)"} & 3 \\ \hline
\end{tabular}%
}
\end{table}

\begin{table}[H]
\centering
\caption{Themes for Question 5}
\label{tab:table-rq5}
\resizebox{\textwidth}{!}{%
\begin{tabular}{|cclc|}
\hline
\multicolumn{4}{|c|}{\textit{\textbf{Q5. How should the suggestion be customized?}}} \\ \hline
\multicolumn{1}{|c|}{\textit{\textbf{Theme}}} & \multicolumn{1}{c|}{\textit{\textbf{Theme frequency}}} & \multicolumn{1}{c|}{\textit{\textbf{Codes}}} & \textit{\textbf{Code frequency}} \\ \hline
\multicolumn{1}{|c|}{\multirow{5}{*}{\textbf{\begin{tabular}[c]{@{}c@{}}Activation timing of\\ suggestions and explanations\end{tabular}}}} & \multicolumn{1}{c|}{\multirow{5}{*}{\textbf{6}}} & \multicolumn{1}{p{7cm}|}{Code 1: "Adjust activation timing manually"} & 2 \\ \cline{3-4} 
\multicolumn{1}{|c|}{} & \multicolumn{1}{c|}{} & \multicolumn{1}{p{7cm}|}{Code 2: "Enable/Disable automatic completion"} & 3 \\ \cline{3-4} 
\multicolumn{1}{|c|}{} & \multicolumn{1}{c|}{} & \multicolumn{1}{p{7cm}|}{Code 3: "Customize activation shortcut"} & 1 \\ \cline{3-4} 
\multicolumn{1}{|c|}{} & \multicolumn{1}{c|}{} & \multicolumn{1}{p{7cm}|}{Code 4: "Customize position of toggle button"} & 1 \\ \cline{3-4} 
\multicolumn{1}{|c|}{} & \multicolumn{1}{c|}{} & \multicolumn{1}{p{7cm}|}{Code 5: Activation of explanations should be customizable by the user} & 2 \\ \hline
\multicolumn{1}{|c|}{\multirow{2}{*}{\textbf{Appearance}}} & \multicolumn{1}{c|}{\multirow{2}{*}{\textbf{6}}} & \multicolumn{1}{p{7cm}|}{Code 1: "Adjust font and style of suggestions"} & 4 \\ \cline{3-4} 
\multicolumn{1}{|c|}{} & \multicolumn{1}{c|}{} & \multicolumn{1}{p{7cm}|}{Code 2: "Define the position of the suggestions"} & 4 \\ \hline
\multicolumn{1}{|c|}{\multirow{3}{*}{\textbf{Coding style}}} & \multicolumn{1}{c|}{\multirow{3}{*}{\textbf{4}}} & \multicolumn{1}{p{7cm}|}{Code 1: "Set the coding style for the generated code"} & 3 \\ \cline{3-4} 
\multicolumn{1}{|c|}{} & \multicolumn{1}{c|}{} & \multicolumn{1}{p{7cm}|}{Code 2: "Set the standards to comply with for the generated code"} & 3 \\ \cline{3-4} 
\multicolumn{1}{|c|}{} & \multicolumn{1}{c|}{} & \multicolumn{1}{p{7cm}|}{Code 3: "Define which data is used for fine tuning the code generation model"} & 2 \\ \hline
\multicolumn{1}{|c|}{\multirow{5}{*}{\textbf{\begin{tabular}[c]{@{}c@{}}Granularity of suggestions\\ and explanations\end{tabular}}}} & \multicolumn{1}{c|}{\multirow{5}{*}{\textbf{6}}} & \multicolumn{1}{p{7cm}|}{Code 1: Customize the explanation granularity} & 2 \\ \cline{3-4} 
\multicolumn{1}{|c|}{} & \multicolumn{1}{c|}{} & \multicolumn{1}{p{7cm}|}{Code 2: "Control the presence of comments in the generated code"} & 2 \\ \cline{3-4} 
\multicolumn{1}{|c|}{} & \multicolumn{1}{c|}{} & \multicolumn{1}{p{7cm}|}{Code 3: "Control amount of code provided by tool"} & 4 \\ \cline{3-4} 
\multicolumn{1}{|c|}{} & \multicolumn{1}{c|}{} & \multicolumn{1}{p{7cm}|}{Code 4: "Number of different suggestions given"} & 1 \\ \cline{3-4} 
\multicolumn{1}{|c|}{} & \multicolumn{1}{c|}{} & \multicolumn{1}{p{7cm}|}{Code 5: "Limit the scope of the code generation"} & 2 \\ \hline
\end{tabular}%
}
\end{table}

\begin{table}[H]
\centering
\caption{Themes for Question 6}
\label{tab:table-rq6}
\resizebox{\textwidth}{!}{%
\begin{tabular}{|cclc|}
\hline
\multicolumn{4}{|c|}{\textit{\textbf{Q6. Where to show the code completion explanation in the UI?}}} \\ \hline
\multicolumn{1}{|c|}{\textit{\textbf{Theme}}} & \multicolumn{1}{c|}{\textit{\textbf{Theme frequency}}} & \multicolumn{1}{c|}{\textit{\textbf{Codes}}} & \textit{\textbf{Code frequency}} \\ \hline
\multicolumn{1}{|c|}{\multirow{4}{*}{\textbf{Inline comments}}} & \multicolumn{1}{c|}{\multirow{4}{*}{\textbf{6}}} & \multicolumn{1}{p{7cm}|}{Code 1: "Comment embedded in the code for context"} & 4 \\ \cline{3-4} 
\multicolumn{1}{|c|}{} & \multicolumn{1}{c|}{} & \multicolumn{1}{p{7cm}|}{Code 2: "Explanation be shown together with the generated code"} & 2 \\ \cline{3-4} 
\multicolumn{1}{|c|}{} & \multicolumn{1}{c|}{} & \multicolumn{1}{p{7cm}|}{Code 3: "Avoid to show complete explanations in comments but only show a brief title"} & 1 \\ \cline{3-4} 
\multicolumn{1}{|c|}{} & \multicolumn{1}{c|}{} & \multicolumn{1}{p{7cm}|}{Code 4: "Display explanation next to the code suggestion"} & 1 \\ \hline
\multicolumn{1}{|c|}{\multirow{7}{*}{\textbf{Separate window}}} & \multicolumn{1}{c|}{\multirow{7}{*}{\textbf{7}}} & \multicolumn{1}{p{7cm}|}{Code 1: "Separate window or tab for explanations"} & 3 \\ \cline{3-4} 
\multicolumn{1}{|c|}{} & \multicolumn{1}{c|}{} & \multicolumn{1}{p{7cm}|}{Code 2: "Define the position of the suggestions"} & 1 \\ \cline{3-4} 
\multicolumn{1}{|c|}{} & \multicolumn{1}{c|}{} & \multicolumn{1}{p{7cm}|}{Code 3: Explanation must be shown separated from the code} & 1 \\ \cline{3-4} 
\multicolumn{1}{|c|}{} & \multicolumn{1}{c|}{} & \multicolumn{1}{p{7cm}|}{Code 4: "Display explanation in a dedicated lateral window"} & 4 \\ \cline{3-4} 
\multicolumn{1}{|c|}{} & \multicolumn{1}{c|}{} & \multicolumn{1}{p{7cm}|}{Code 5: "Show detailed explanation in a separated window"} & 1 \\ \cline{3-4} 
\multicolumn{1}{|c|}{} & \multicolumn{1}{c|}{} & \multicolumn{1}{p{7cm}|}{Code 6: "Provide explanation in a distinct popup"} & 2 \\ \cline{3-4} 
\multicolumn{1}{|c|}{} & \multicolumn{1}{c|}{} & \multicolumn{1}{p{7cm}|}{Code 7: "Sidebar tab to show explanations alongside code completion"} & 3 \\ \hline
\multicolumn{1}{|c|}{\textbf{Chatbot}} & \multicolumn{1}{c|}{\textbf{3}} & \multicolumn{1}{p{7cm}|}{Code 1: "Explanations can be asked to a chatbot"} & 3 \\ \hline
\end{tabular}%
}
\end{table}

\begin{table}[H]
\centering
\caption{Themes for Question 7}
\label{tab:table-rq7}
\resizebox{\textwidth}{!}{%
\begin{tabular}{|cclc|}
\hline
\multicolumn{4}{|c|}{\textit{\textbf{Q7. When to show the explanation during the interaction?}}} \\ \hline
\multicolumn{1}{|c|}{\textit{\textbf{Theme}}} & \multicolumn{1}{c|}{\textit{\textbf{Theme frequency}}} & \multicolumn{1}{c|}{\textit{\textbf{Codes}}} & \textit{\textbf{Code frequency}} \\ \hline
\multicolumn{1}{|c|}{\multirow{4}{*}{\textbf{\begin{tabular}[c]{@{}c@{}}Manually interacting with\\ the suggested code\end{tabular}}}} & \multicolumn{1}{c|}{\multirow{4}{*}{\textbf{6}}} & \multicolumn{1}{p{7cm}|}{Code 1: "Explanations should be provided for pieces of code selected by the user"} & 3 \\ \cline{3-4} 
\multicolumn{1}{|c|}{} & \multicolumn{1}{c|}{} & \multicolumn{1}{p{7cm}|}{Code 2: "Explanations appear when clicking the mouse"} & 2 \\ \cline{3-4} 
\multicolumn{1}{|c|}{} & \multicolumn{1}{c|}{} & \multicolumn{1}{p{7cm}|}{Code 3: "Explanations appear when hovering over generated code"} & 2 \\ \cline{3-4} 
\multicolumn{1}{|c|}{} & \multicolumn{1}{c|}{} & \multicolumn{1}{p{7cm}|}{Code 4: "Explanations should be manually expandable"} & 3 \\ \hline
\multicolumn{1}{|c|}{\textbf{Shortcut}} & \multicolumn{1}{c|}{\textbf{3}} & \multicolumn{1}{p{7cm}|}{Code 2: "Explanations are invoked by clicking a key or combination of keys"} & 3 \\ \hline
\multicolumn{1}{|c|}{\multirow{3}{*}{\textbf{\begin{tabular}[c]{@{}c@{}}Automatic short explanation,\\ details on-demand\end{tabular}}}} & \multicolumn{1}{c|}{\multirow{3}{*}{\textbf{3}}} & \multicolumn{1}{p{7cm}|}{Code 1: "Automatic explanation for short suggestions"} & 3 \\ \cline{3-4} 
\multicolumn{1}{|c|}{} & \multicolumn{1}{c|}{} & \multicolumn{1}{p{7cm}|}{Code 2: "Explanation shown automatically to non-expert users"} & 1 \\ \cline{3-4} 
\multicolumn{1}{|c|}{} & \multicolumn{1}{c|}{} & \multicolumn{1}{p{7cm}|}{Code 3: "Long explanations should be provided only on demand"} & 1 \\ \hline
\end{tabular}%
}
\end{table}

\begin{table}[H]
\centering
\caption{Themes for Question 8}
\label{tab:table-rq8}
\resizebox{\textwidth}{!}{%
\begin{tabular}{|cclc|}
\hline
\multicolumn{4}{|c|}{\textit{\textbf{Q8. What should be explained?}}} \\ \hline
\multicolumn{1}{|c|}{\textit{\textbf{Theme}}} & \multicolumn{1}{c|}{\textit{\textbf{Theme frequency}}} & \multicolumn{1}{c|}{\textit{\textbf{Codes}}} & \textit{\textbf{Code frequency}} \\ \hline
\multicolumn{1}{|c|}{\multirow{3}{*}{\textbf{Purpose}}} & \multicolumn{1}{c|}{\multirow{3}{*}{\textbf{5}}} & \multicolumn{1}{p{7cm}|}{Code 1: "Explain why the code was generated in a specific manner"} & 4 \\ \cline{3-4} 
\multicolumn{1}{|c|}{} & \multicolumn{1}{c|}{} & \multicolumn{1}{p{7cm}|}{Code 2: "List pros and cons about the generated code"} & 1 \\ \cline{3-4} 
\multicolumn{1}{|c|}{} & \multicolumn{1}{c|}{} & \multicolumn{1}{p{7cm}|}{Code 3: "Explain different possible alternatives"} & 1 \\ \hline
\multicolumn{1}{|c|}{\multirow{3}{*}{\textbf{Examples and references}}} & \multicolumn{1}{c|}{\multirow{3}{*}{\textbf{5}}} & \multicolumn{1}{p{7cm}|}{Code 1: "Provide real-world examples for better understanding"} & 4 \\ \cline{3-4} 
\multicolumn{1}{|c|}{} & \multicolumn{1}{c|}{} & \multicolumn{1}{p{7cm}|}{Code 2: "Add documentation references"} & 3 \\ \cline{3-4} 
\multicolumn{1}{|c|}{} & \multicolumn{1}{c|}{} & \multicolumn{1}{p{7cm}|}{Code 3: "Use trusted sources to report additional information in the explanation"} & 2 \\ \hline
\multicolumn{1}{|c|}{\textbf{Contextual explanation}} & \multicolumn{1}{c|}{\textbf{4}} & \multicolumn{1}{p{7cm}|}{Code 1: "Provide suggestion of the selected piece of code"} & 4 \\ \hline
\multicolumn{1}{|c|}{\multirow{3}{*}{\textbf{\begin{tabular}[c]{@{}c@{}}Minimal explanation,\\ details on-demand\end{tabular}}}} & \multicolumn{1}{c|}{\multirow{3}{*}{\textbf{4}}} & \multicolumn{1}{p{7cm}|}{Code 1: "Explanation should be minimal, with the possibility to manually ask for more detail"} & 2 \\ \cline{3-4} 
\multicolumn{1}{|c|}{} & \multicolumn{1}{c|}{} & \multicolumn{1}{p{7cm}|}{Code 2: "It would be useful to provide the explanation as a dialogue in a chat interaction"} & 1 \\ \cline{3-4} 
\multicolumn{1}{|c|}{} & \multicolumn{1}{c|}{} & \multicolumn{1}{p{7cm}|}{Code 3: "Provide an explanation preview that is expandable onDemand"} & 1 \\ \hline
\multicolumn{1}{|c|}{\multirow{5}{*}{\textbf{Functionality}}} & \multicolumn{1}{c|}{\multirow{5}{*}{\textbf{4}}} & \multicolumn{1}{p{7cm}|}{Code 1: "Explanation should describe what the code is and does"} & 3 \\ \cline{3-4} 
\multicolumn{1}{|c|}{} & \multicolumn{1}{c|}{} & \multicolumn{1}{p{7cm}|}{Code 2: "Explanation should cover what data structures and code constructs appear in the generated code"} & 2 \\ \cline{3-4} 
\multicolumn{1}{|c|}{} & \multicolumn{1}{c|}{} & \multicolumn{1}{p{7cm}|}{Code 3: "Show pseudocode or flowcharts"} & 2 \\ \cline{3-4} 
\multicolumn{1}{|c|}{} & \multicolumn{1}{c|}{} & \multicolumn{1}{p{7cm}|}{Code 4: "Explanation should cover aspects that are not familiar to the user"} & 1 \\ \cline{3-4} 
\multicolumn{1}{|c|}{} & \multicolumn{1}{c|}{} & \multicolumn{1}{p{7cm}|}{Code 5: "Explanation should also explain low level concepts such as syntax and data types"} & 1 \\ \hline
\end{tabular}%
}
\end{table}